\documentclass[11pt]{article}
\pdfoutput=1
\usepackage[margin=1in]{geometry}
\usepackage{latexsym}
\usepackage{mathptmx}
\usepackage{amsfonts,amsthm,amssymb,amsmath}
\usepackage{graphicx}
\usepackage[dvipsnames,usenames]{color}
\usepackage{xspace}
\usepackage{caption}
\usepackage{subcaption}
\usepackage{framed}

\usepackage{enumitem}
\usepackage{bm}
\usepackage{multirow}
\usepackage{graphicx}
\usepackage[boxed]{algorithm}
\usepackage[noend]{algorithmic}
\usepackage[sort]{cite}
\usepackage{typearea}
\typearea{13}

\usepackage[compact]{titlesec}

\usepackage[utf8]{inputenc}
\usepackage[english]{babel}

\newcommand{\osd}{{\sc osd}\xspace}

\newcommand{\child}{{\tt child}\xspace}

\newcommand{\opt}{{\tt opt}\xspace}
\newcommand{\alg}{{\tt alg}\xspace}

\newcommand{\ps}{{\sc ps}\xspace}

\newcommand{\ex}{{\mathbb E}}

\newcommand{\eat}[1]{}

\newcounter{note}[section]
\newcommand{\debmalya}[1]{\refstepcounter{note}$\ll${\sf Debmalya's Comment~\thenote:} {\sf \textcolor{red}{#1}}$\gg$\marginpar{\tiny\bf DP~\thenote}}

\newcommand{\mcm}{{\sc mcm}\xspace}

\newtheorem{theorem}{Theorem}
\newtheorem{lemma}[theorem]{Lemma}

\title{Online Service with Delay}
\author{
Yossi Azar\thanks{Blavatnik School of Computer Science, Tel Aviv University. Email: {\tt azar@tau.ac.il}.}
\and 
Arun Ganesh\thanks{Department of Computer Science, Duke University. Email: {\tt arun.ganesh@duke.edu}.} 
\and 
Rong Ge\thanks{Department of Computer Science, Duke University. Email: {\tt rongge@cs.duke.edu}.} 
\and 
Debmalya Panigrahi\thanks{Department of Computer Science, Duke University. Email: {\tt debmalya@cs.duke.edu}.}
}
\date{}

\begin{document}

\maketitle

\thispagestyle{empty}

\begin{abstract}
In this paper, we introduce the {\em online service with delay} problem.
In this problem, there are $n$ points in a metric space that issue service requests 
over time, and a server that serves these requests. The goal is to minimize the
sum of distance traveled by the server and the total delay in serving the requests.
This problem models the fundamental
tradeoff between batching requests to improve locality and reducing delay 
to improve response time, that has many applications
in operations management, operating systems, logistics, 
supply chain management, and scheduling.

Our main result is to show a poly-logarithmic competitive ratio for the online service with delay
problem. This result is obtained by an algorithm that we call the 
{\em preemptive service algorithm}. The salient feature of this algorithm
is a process called preemptive service, which uses a novel combination 
of (recursive) time forwarding and spatial exploration on a metric space. We hope this
technique will be useful for related problems such as reordering buffer management, 
online TSP, vehicle routing, etc. We also generalize our results to $k > 1$ servers.
\end{abstract}

\clearpage

\setcounter{page}{1}

\section{Introduction}
\label{sec:introduction}

Suppose there are $n$ points in a metric space that issue service requests 
over time, and a server that serves these requests. A request can be served
at any time after it is issued. The goal is to minimize
the sum of distance traveled by the server ({\em service cost}) and the total
delay of serving requests ({\em delay} of a request is the difference between
the times when the request is issued and served). We call this the {\em online service
with delay} (\osd) problem. To broaden the scope of the problem, for each request we also
allow any {\em delay penalty} that is a non-negative monotone
function of the actual
delay, the penalty function being revealed when the request is issued. 
Then, the goal is to minimize the sum of service cost and delay penalties.

This problem captures many natural scenarios. For instance, consider the 
problem of determining the schedule of a technician attending service calls 
in an area. It is natural to prioritize service in areas that have a large
number of calls, thereby balancing the penalties incurred for delayed service 
of other requests with the actual cost of dispatching the technician.
In this context, the delay penalty function can be loosely interpreted as the 
level of criticality of a request, and different requests, even at the same 
location, can have very different penalty functions. In 
general, the \osd problem formalizes the tradeoff between batching requests
to minimize service cost by exploiting locality and quick response 
to minimize delay or response time. Optimizing this tradeoff is a fundamental
problem in many areas of operations management, operating systems, logistics, 
supply chain management, and scheduling.

A well-studied problem with a similar motivation is the reordering buffer 
management problem~\cite{RackeSW02,EnglertW05,KhandekarP06,EnglertRW07,GamzuS09,
Avigdor-ElgrabliR10,AdamaszekCER11,BarmanCU12,Avigdor-ElgrabliR13,
Avigdor-ElgrabliIMR15} --- the only difference with the \osd problem 
is that instead of 
delay penalties, there is a cap on the number of unserved requests at any time. 
Thus, the objective only contains the service cost, and the delays appear
in the constraint. A contrasting class of problems are the online traveling 
salesman problem and its many variants~\cite{irani2004line,krumke2004whack, 
AzarV15,AusielloFLST95,AusielloFLST01},
where the objective is only defined on the delay (average/maximum completion time, 
number of requests serviced within a deadline, and so on), but the server's movement 
is constrained by a given speed (which implies that there are no competitive algorithms 
for delay in general, so these results restrict the sequence or the adversary in
some manner).
In contrast to these problem classes, both the
service cost and the delay penalties appear in the objective of the \osd
problem. In this respect, the \osd problem bears similarities to 
the online multi-level aggregation problem~\cite{KhannaNR02,BienkowskiBBCDFJSTV16,BuchbinderFNT17}, where a server
residing at the root of a tree serves requests arriving online at the 
leaves after aggregating them optimally. 
While these problems are incomparable from a technical perspective, 
all of them represent natural abstractions of the fundamental tradeoff that
we noted above, 
and the right abstraction depends on the specific application.

Our main result is a a poly-logarithmic competitive ratio for the  
\osd problem in general metric spaces.  

\begin{theorem}
\label{thm:osd}
	There is a randomized algorithm with a competitive ratio of $O(\log^4 n)$
    for the \osd problem.
\end{theorem}

Before proceeding further, let us try to understand why the \osd problem
is technically interesting. Recall that we wish to balance service costs with 
delay penalties. Consider the following natural algorithm. Let us represent 
the penalty for all unserved requests at a location as a ``ball'' growing out 
of this location. These balls collide with each other and merge, and grow 
further together, until they reach the server's current location. At this 
point, the server moves and serves all requests whose penalty balls reached it.
Indeed, this algorithm achieves the desired balance --- the total delay penalty
is equal (up to constant factors) to the total service cost. But, is this algorithm 
competitive against the optimal solution? Perhaps surprisingly, it is not! 

\begin{figure}
	\centering
    \includegraphics[width=0.75\columnwidth]{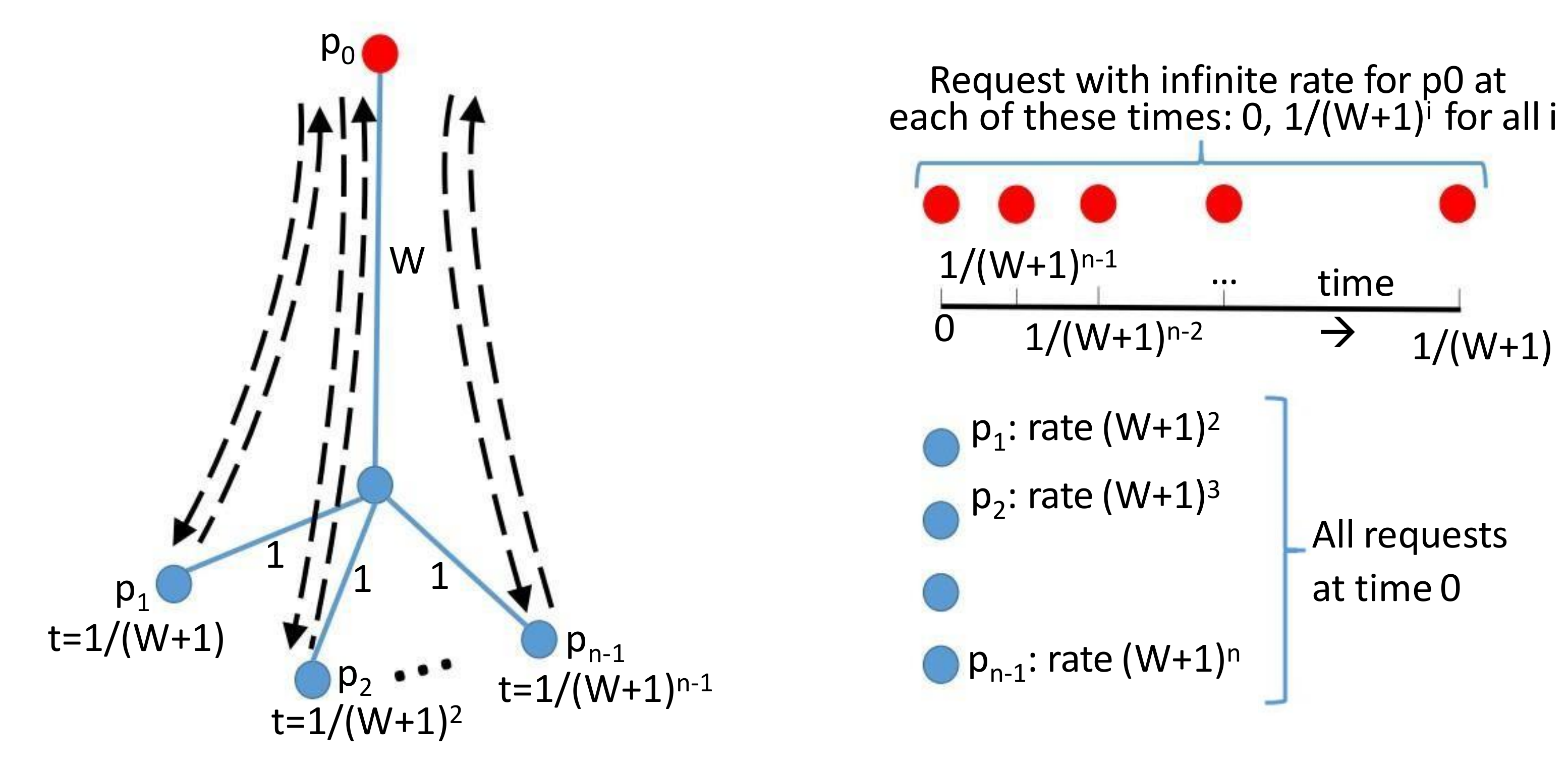} 
	\caption{\small An example showing the inadequacy of simple balancing of 
    delay penalties with service costs.}
	\label{fig:preempt}
\end{figure}

Consider the instance in Fig.~\ref{fig:preempt} on a star metric, where 
location $p_0$ is connected to the center with an edge of length $W$ ($\gg 1$), 
but all other locations $p_1, p_2, \ldots, p_{n-1}$ are connected using edges of 
length 1. The delay penalty function for each request is a constant 
request-specific rate times the delay.
All requests for locations $p_i$, $i \geq 1$ arrive at time $0$, with
$p_i$ getting a single request accumulating waiting cost at rate $(W+1)^{i+1}$. 
Location $p_0$ is special in 
that it gets a request with an infinite\footnote{or sufficiently large} rate at time $0$ and at all times $1/(W+1)^i$. For this instance, 
the algorithm's server will move to location $p_i$ and back to $p_0$ at each 
time $1/(W+1)^i$. This is because the ``delay ball'' for $p_i$ reaches 
$p_0$ at time $1/(W+1)^i$, and the ``delay ball'' for $p_0$ instantly 
reaches $p_i$ immediately afterward. Note, however, that the ``delay balls'' of all
locations $p_j$, $j < i$, have not crossed their individual edges connecting
them to the center at this time. Thus, the algorithm incurs a total cost of
$\Omega(nW)$. On the other hand, 
the optimal solution serves all the requests for $p_1, p_2, \ldots, p_{n-1}$ 
at time 0, moves to location $p_0$, and stays there the entire time. Then, 
the optimal solution incurs a total cost of only $O(n+W)$.

\begin{figure}
	\centering
    \includegraphics[width=0.6\columnwidth]{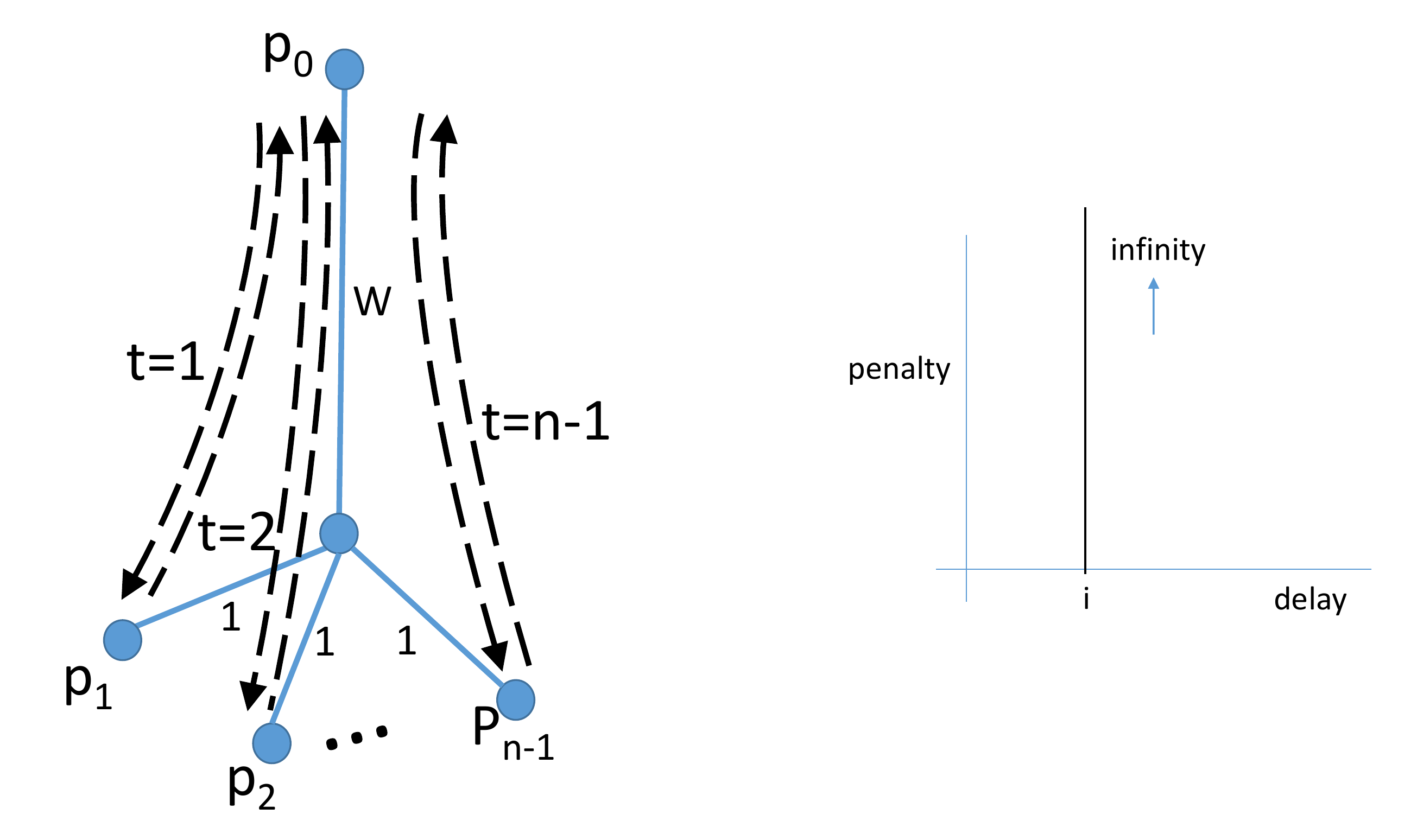}
	\caption{\small An example in which requests must be served preemptively
    even before incurring any delay penalty. The function on the right is the 
    delay penalty function. The penalty is $0$ till delay $i$, at which point 
    the penalty goes to infinity. Page $p_i$, for each $i \geq 1$, has 
    a single request arriving at time $0$. Page $p_0$ has requests at times
    $0, 1, 2, \ldots, n-1$.}
	\label{fig:preempt-deadline}
\end{figure}

In this example, the algorithm must serve requests at $p_1$, $p_2$, $\ldots$, $p_n$ 
even when they have incurred a very small delay penalty. In fact, this example
can be modified to create situations where requests that have not incurred any
delay penalty at all need to be served! 
Consider the instance in Fig.~\ref{fig:preempt-deadline} on the same star metric with different delay penalties.
Note that the delay penalty enforces
deadlines -- a request for $p_i$ must incur delay 
$\leq i$. (Requests for $p_0$ must be served immediately.) 
On this instance, the ``ball'' growing algorithm 
moves to location $p_i$ and back to $p_0$ for each
time $i \geq 0$, incurring a total cost of $\Omega(nW)$. On the other hand, 
the optimal solution serves all the requests for $p_1, p_2, \ldots, p_{n-1}$ 
immediately after serving the request for $p_0$ at time 0, and then 
stays at location $p_0$ for the entire remaining time, thereby incurring a
total cost of only $O(n+W)$. In order to be competitive
on this example, an \osd algorithm must serve requests at 
$p_1, p_2, \ldots, p_n$ 
even before they have incurred any delay penalty.

This illustrates an important requirement of an \osd algorithm --- although
it is trying to balance delay penalties and service costs, it cannot hope to 
do so ``locally'' for the set of requests at any single location
(or in nearby locations, if their penalty balls have merged). Instead, it must
perform ``global'' balancing of the two costs, since it has to serve requests
that have not accumulated any delay penalty. This principle, that we call 
{\em preemptive service}, is what makes the \osd problem both interesting
and challenging from a technical perspective, and the main novelty of our
algorithmic contribution will be in addressing this requirement. Indeed,
we will call our algorithm the {\em preemptive service algorithm} 
(or \ps algorithm).

As we will discuss later, a natural choice for preemptive service is to serve requests which will go critical the earliest. However, on more general metrics than a star, preemptive service must also consider the spatial locality of non-critical requests.

\begin{figure}
	\centering
    \includegraphics[width=0.5\columnwidth]{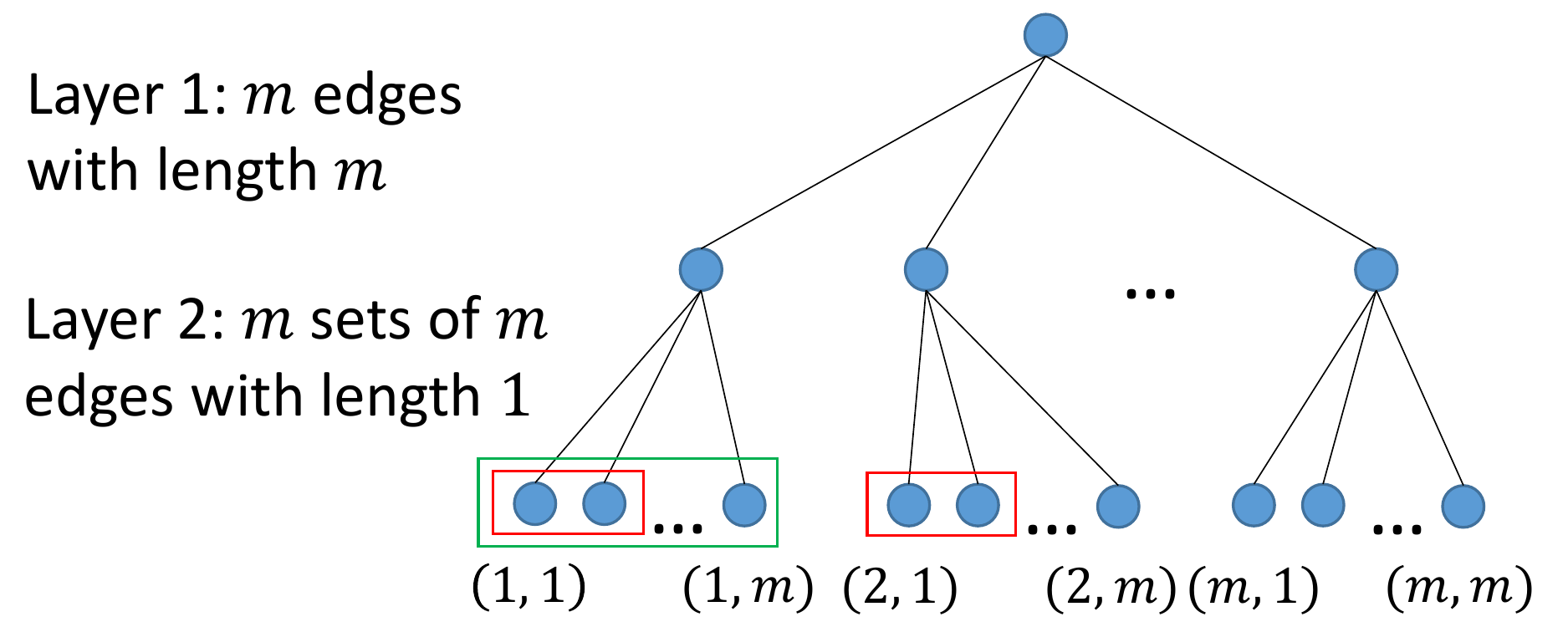}
	\caption{\small An example showing that spatial locality should be considered in preemptive service. The red boxes show a set of requests that might be served using budget $2m$ if spatial locality is not considered. The green box shows a set of requests that might be served using budget $2m$ if spatial locality is considered.}
	\label{fig:spatial-locality}
\end{figure}

Consider a subtree where we would like to perform preemptive service (see Fig. \ref{fig:spatial-locality}). The subtree has two layers, with $m$ edges of length $m$ in the first layer. Each of these $m$ edges also has $m$ children that are connected by edge of length $1$. We use the pair $(i,j)$ to denote the $j$-th children of the $i$-th subtree. Each of these nodes have a request. Now suppose the ordering for these requests to become critical is $(1,1)$, $(2,1)$, $\dots$, $(m,1)$, $(1, 2)$, $(2,2)$, $\dots$, $(m,2)$, $\dots$, $(m,m)$. If we have a budget of $2cm$ for the preemptive service for some $c < m$, we can only serve at most $2c$ requests if we follow the order in which they will become critical. The total cost for the entire tree would be $\Theta(m^2)$. However we can choose to preemptively serve all requests in $c$ of the subtrees, and in this way we can finish preempting the whole tree for $O(m)$ cost.



\smallskip\noindent{\bf \osd on HSTs.}
Our main theorem (Theorem~\ref{thm:osd}) is obtained as a corollary of 
our following result on an HST.
A {\em hierarchically separated tree} (HST) is a rooted tree where every 
edge length is shorter by a factor of at least $2$ from its parent edge 
length.\footnote{In general, the ratio of lengths of a parent and 
a child edge is a parameter of the HST and 
need not be equal to $2$ . Our results for HSTs 
also extend to other constants bounded away from $1$, but for simplicity, 
we will fix this parameter to $2$ in this paper. Furthermore, note that we can round all edge lengths down to the nearest length of the form $2^i$ for some integer $i$ and obtain power-of-$2$ ratios between all edge lengths, while only distorting distances in the HST by at most a factor of $2$.}  
\begin{theorem}
\label{thm:osd-hst}
	There is a deterministic algorithm with a competitive ratio of $O(h^3)$
    for the \osd problem on a hierarchically separated tree of depth $h$.
\end{theorem}
Theorem~\ref{thm:osd} follows from Theorem~\ref{thm:osd-hst} by a 
standard probabilistic embedding of general metric spaces in HSTs of depth 
$O(\log n)$ with an expected distortion of $O(\log n)$ in 
distances:

\begin{theorem}
\label{thm:embedding}
For any $n$-point metric space $\cal M$, there exists a distribution
$\cal D$ over HSTs of depth $O(\log n)$, such that the expected 
distortion is $O(\log n)$. In other words, for any two points $(u, v)$, 
the following holds:
\begin{equation*}
	d_{\cal M} (u, v) \leq \ex_{T \sim {\cal D}} [d_T (u, v)] \leq O(\log n) d_{\cal M} (u, v),
\end{equation*}
where $d_{\cal M} (u, v)$ and $d_T (u, v)$ denote the distance
between $u$ and $v$ in the metric space $\cal M$ and the HST $T$ respectively.
The $n$ points in the metric space $\cal M$ are leaves in each of the HSTs in 
the distribution.
\end{theorem}

This is done in two steps. Define a $2$-HST as a rooted tree where every edge length is shorter by a factor 
of at exactly $2$ from its parent edge length. First, a well-known result of Fakcharoenphol~{\em et al.} \cite{FakcharoenpholRT04} states that any metric space can be embedded into a distribution over $2$-HSTs with distortion $O(\log n)$, where the points of the metric space are the leaves of the trees. The depth of 
the HSTs is, however, dependent on the aspect ratio of the metric space, i.e., the maximum ratio between distances of two pairs of points in $\cal M$. On the other hand, a result of Bansal~{\em et al.} \cite{BansalBMN15} (Theorem 8 of the paper) states that a $2$-HST with $n$ leaves can be deterministically embedded into an HST of height $O(\log n)$ with the same leaf set, such that the distance between any pair of leaves is distorted by at most $4$. The theorem follows by composing these two embeddings.

Note that Theorem~\ref{thm:osd-hst} also implies 
$O(1)$ deterministic algorithms for the \osd problem on the interesting
special cases of the uniform metric 
and any star metric, since these are depth-1 HSTs. 


\smallskip\noindent
{\bf Generalization to $k$ servers.} 
We also generalize the \osd problem to $k > 1$ servers. This problem
generalizes the well-known online paging (uniform metric),
weighted paging (star metric), and $k$-server (general metric)
problems. We obtain an analogous theorem to Theorem~\ref{thm:osd-hst} on HSTs,
which again extends to an analogous theorem to Theorem~\ref{thm:osd}
on general metrics. 

\begin{theorem}
\label{thm:kosd}
	There is a deterministic algorithm with a competitive ratio of $O(k h^4)$
    for the $k$-\osd problem on an HST of depth $h$.
    As immediate corollaries, this yields the following:
    \begin{itemize}
		\item A randomized algorithm with a competitive ratio of $O(k\log^5 n)$
	    for the $k$-\osd problem on general metric spaces.
        \item A deterministic algorithm with a competitive ratio of 
        $O(k)$ for the unweighted and weighted paging problems with delay 
        penalties, which are respectively special cases of $k$-\osd on 
        the uniform metric and star metrics.
   \end{itemize}
\end{theorem}

Because of the connections to paging and $k$-server,
the competitive ratio for the $k$-\osd problem is worse by a factor of $k$
compared to that for the \osd problem. The algorithm for $k$-\osd requires
several new ideas specific to $k>1$. We give details of these results and techniques in Section~\ref{sec:kosd}.

\smallskip\noindent
{\bf Non-clairvoyant \osd.}
One can also consider the {\em non-clairvoyant} version of the 
\osd problem, where the algorithm only knows the current delay penalty
of a request but not the entire delay penalty function.
Interestingly, it turns out that there is a fundamental distinction 
between the uniform metric space and general metric spaces in this case.
The results for \osd in uniform metrics carry over to the 
non-clairvoyant version.
In sharp contrast, we show a lower bound of $\Omega(\Delta)$,
where $\Delta$ is the aspect ratio,\footnote{{\em Aspect ratio} is the ratio 
of maximum to minimum distance between pairs of points in a metric space.} 
even for a star metric in the non-clairvoyant setting.
Our results for the uniform and star metrics appear in Section~\ref{sec:special}.

\smallskip\noindent
{\bf Open Problems.} 
The main open question that arises from our work is whether there exists an 
$O(1)$-competitive algorithm for the \osd problem. This would require
a fundamentally different approach, since the embedding into an HST already loses
a logarithmic factor. Another interesting question is to
design a randomized algorithm for $k > 1$ servers that has a logarithmic
dependence on $k$ in the competitive ratio. 
We show this for the uniform metric, and leave 
the question open for general metrics. Even for
a star metric, the only known approach for the classical 
$k$-server problem (which is a special case) is via an LP relaxation, and 
obtaining an LP relaxation for the \osd problem seems challenging.
Finally, one can also consider the {\em offline} version of 
this problem, where the release times of requests are known
in advance.

In the rest of this section, we outline the main techniques that we 
use in the \ps algorithm. Then, we describe the \ps algorithm 
on an HST in Section~\ref{sec:single-server-algo}, and prove its 
competitive ratio in Section~\ref{sec:single-server-analysis}. In Section~\ref{sec:kosd} we generalize these results to $k$-\osd. 
Section~\ref{sec:special} contains our results on \osd and $k$-\osd
for the uniform metric and star metrics, which correspond to 
(weighted) paging problems with delay.

\subsection{Our Techniques}

Our algorithmic ideas will be used for \osd on general HSTs, 
but we initially describe them 
on a star metric for simplicity. 

\smallskip \noindent {\bf \osd on a star metric.}
Recall our basic algorithmic strategy of trying
to equalize delay penalties with service cost. For this purpose, we outlined a ball
growing process earlier. To implement this process, 
let us place a counter on every edge.
We maintain the invariant that every unserved request increments one of the  
counters on its path to the server by its instantaneous delay penalty at all times. 
Once the value of an edge counter reaches the length of the edge, it is said to 
be {\em saturated} and cannot increase further. We call a request {\em critical}
if all the counters on edges connecting it to the server are saturated. The 
algorithm we outlined earlier moves the server whenever there is any critical
request, and serves all critical requests. 
For every edge that the server traverses, its counter is reset since all requests 
increasing its counter are critical and will be served.
As we observed, this algorithm has the property that the cost of serving 
critical requests equals (up to a constant) the total delay penalty of those
requests. 

\medskip\noindent{\bf Preemptive Service.}
But, as we saw earlier, this algorithm is not competitive. 
Let us return to the example in Fig.~\ref{fig:preempt}.
In this example, the algorithm needs to decide when to serve the requests for 
$p_1,p_2,...,p_{n-1}$. Recall that the algorithm fails if it waits until
these requests have accumulated sufficiently large delay penalties. 
Instead, it must {\em preemptively} serve requests before their delay penalty 
becomes large. This implies that the algorithm must perform two different
kinds of service: {\em critical service} for critical requests, and 
{\em preemptive service} for non-critical requests. For preemptive service,
the algorithm needs to decide two things: {\em when} to perform preemptive service and 
{\em which} unserved requests to preemptively serve. For the first question, we use a 
simple piggybacking strategy: whenever the algorithm decides to perform critical
service, it might as well serve other non-critical 
requests preemptively whose overall service cost is similar to that of 
critical service. This ensures that preemptive 
service is essentially ``free'', i.e., can be charged to critical service
just as delay penalties were being charged.

\smallskip \noindent {\bf Time Forwarding.} 
But, how does the algorithm prioritize between non-critical requests 
for preemptive service? 
Indeed, if future delay penalties are unknown, there is no way for the 
algorithm to prioritize correctly between requests for pages 
$p_1, p_2, \ldots, p_{n-1}$ in Fig.~\ref{fig:preempt}. 
(This is what we use in our lower bound for the non-clairvoyant setting,
where the algorithm does not know future delay penalties.) 
\eat{
Indeed, we show that this example can be modified to give a lower bound for the
weighted paging problem in the non-clairvoyant setting (proof in appendix).
\begin{theorem}
\label{thm:non-clairvoyant}
	There is a lower bound of $\Omega(W)$ on the competitive ratio of weighted paging
	with delay in the non-clairvoyant setting, where $W$ is the ratio of the maximum 
    to minimum page weight.
\end{theorem}

In the clairvoyant setting, however, the algorithm has more information: in the instance
in Fig.~\ref{fig:preempt}, 
it exactly knows the identity of the pages $p_1, p_2, \ldots, p_{n-1}$. 
}
This implies that the algorithm must simulate future time in order
to prioritize between the requests. A natural prioritization order would be: 
requests that are going to become critical in the nearest
future should have a higher priority of being served preemptively. One must note that 
this order is not necessarily the order in which requests will {\em actually} become 
critical in the future, since the actual order also depends on future requests.
Nevertheless, we observe that if future requests caused a request to become critical 
earlier, an optimal solution must also either suffer a delay penalty for these
requests or serve this location again. Therefore, the optimal solution's advantage 
over the algorithm is not in future requests, but in previous requests that it has
already served. 
In general, whenever
the algorithm performs service of critical requests, it also simulates the future by 
a process that we call {\em time forwarding} and identifies a subset of requests 
for preemptive service. We will see that this algorithm has a constant competitive 
ratio for \osd on a star metric.

\eat{
\begin{theorem}
\label{thm:paging-clairvoyant}
	There is a deterministic algorithm for weighted paging with delay that has a competitive
	ratio of $O(k)$ for the clairvoyant setting.
\end{theorem}
}
\smallskip \noindent {\bf \osd on an HST.}
How do we extend this algorithm to an HST of arbitrary depth? 
%
%
For critical service, we essentially use the same process as before, placing 
a counter on each edge and increasing them by delay penalties. Some additional
complications are caused by the fact that the server might be at a different 
level in the HST from a request, which requires us to only allow a request
to increase counters on a subset of edges on its path to the server. For 
details of this process, the reader is referred to Section~\ref{sec:single-server-algo}.

\smallskip \noindent {\bf Recursive Time Forwarding.} 
The main difficulty, however, is again in implementing preemptive service, and in particular 
prioritizing between requests for preemption. 
%
%
%
Unlike on a star, 
in a tree, we must balance two considerations: (a) as in a star, 
requests that are going to be 
critical earliest in the future should be preferentially served, but (b) requests
that are close to each other in the tree, i.e., can be served together at low cost,
should also be preferred over requests that are far from each other. 
(Indeed, we have shown an example in Fig. \ref{fig:spatial-locality} where only using criterion (a) fails.) 
To balance these considerations, we devise a 
recursive time forwarding process that identifies a sequence of layers of edges,
forwarding time independently on each of them to identify the next layer. This 
recursive process leads to other complications, such as the fact that rounding the
length of edges in each layer can lead to either an exponential blow up in the 
cost of preemptive service, or an exponential decay in the total volume of 
requests served, both of which are unacceptable. We show how to overcome these 
difficulties  by careful modifications of the algorithm and analysis 
in Sections~\ref{sec:single-server-algo} and 
\ref{sec:single-server-analysis}, eventually obtaining Theorem~\ref{thm:osd-hst}.
Theorem~\ref{thm:osd} now follows by standard techniques, as discussed above.
\eat{
\begin{theorem}
\label{thm:hst}
	There is a deterministic $O(kh)$-competitive algorithm for the $k$-server
	problem with delay on an HST of depth $h$.
\end{theorem}
Combining this theorem with a standard probabilistic embedding of general metric
spaces into a distribution of HSTs~\cite{FakcharoenpholRT04}, 
we obtain the following corollary.

\begin{theorem}
\label{thm:metric}
	There is a randomized $O(k \log^2 n)$-competitive algorithm for the $k$-server
	problem with delay on an $n$-point metric space.
\end{theorem}
}
\subsection{Related Work}

{\bf Reordering Buffer Management.} The difference between reordering buffer management
and the \osd problem is that instead of delay penalties, the 
number of unserved requests cannot exceed a given number $b$ at any time.
%
%
This problem was introduced by R\"{a}cke {\em et al.}~\cite{RackeSW02}, 
who devised an $O(\log^{2}b)$ -competitive algorithm on the uniform 
metric space. The competitive ratio has progressively improved
to $O(\sqrt{\log b})$ for deterministic algorithms and 
$\Omega(\log\log b)$ for randomized algorithms~\cite{EnglertW05, 
Avigdor-ElgrabliR10, AdamaszekCER11, Avigdor-ElgrabliR13}.
The randomized bound is known to be tight, while the best
known deterministic lower bound is 
$\Omega(\sqrt{\log b/\log\log b})$~\cite{AdamaszekCER11}.
This problem has also been studied on 
other metric spaces~\cite{BarmanCU12,KhandekarP06,GamzuS09}. In particular,
Englert {\em et al.}~\cite{EnglertRW07} obtained a competitive
ratio of $O(\log^{2}b \log n)$ for general metric spaces. 
For a star metric, the best deterministic upper bound is 
$O(\sqrt{\log b})$ ~\cite{AdamaszekCER11} and the best 
randomized upper bound is $O(\log\log b\gamma)$, 
where $\gamma$ is the aspect ratio~\cite{Avigdor-ElgrabliIMR15}.
Interestingly, the online algorithm for reordering buffer in general metric spaces 
uses the ball growing approach that we outlined earlier, and serves a request when 
its ``ball'' reaches the server. As we saw, this strategy is not competitive
for \osd, necessitating preemptive service.


\smallskip\noindent
{\bf Online Traveling Salesman and Related Problems.} 
In this class of problems, the server is restricted to move 
at a given speed on the metric space, and the objective is only defined on
the delay in service, typically the goal being to maximize the number 
of requests served within given deadlines.
%
There is a large amount of work for this problem when the
request sequence is known in advance~\cite{bansal2004approximation, 
chekuri2007approximation, bar2005approximating, tsitsiklis1992special,
karuno20032}, and for a special case called orienteering when all 
deadlines are identical and all requests arrive at the 
outset~\cite{arkin1998resource, awerbuch1998new, 
golden1987orienteering, bansal2004approximation, chekuri2012improved,
chekuri2004maximum, blum2003approximation}.
When the request sequence is not known in advance, the 
problem is called dynamic traveling repairman and has been 
considered for various special cases~\cite{irani2004line,
krumke2004whack, AzarV15}. A minimization
version in which there are no deadlines and the goal
is to minimize the maximum/average completion times of the service
time has also been studied~\cite{AusielloFLST95,AusielloFLST01}.
The methods used for these problems are conceptually different from 
our methods since service is limited by server speed and 
the distance traveled is not counted in the objective.


\smallskip
\noindent {\bf Online Multi-level Aggregation.}
In the online multi-level aggregation problem, the input is a tree 
and requests arrive at its leaves. The requests accrue delay penalties and the 
service cost is the total length of edges in the subtree connecting the
requests that are served with the root. In the \osd terminology, this 
corresponds to a server that always resides at the root of the tree. 
This problem was studied by Khanna {\em et al.}~\cite{KhannaNR02}.
More recently, the deadline version, where each request must be served
by a stipulated time but does not accrue any delay penalty,
was studied by Bienkowski {\em et al.}~\cite{BienkowskiBBCDFJSTV16},
who obtained a constant-competitive algorithm for a tree of constrant 
depth. This result was further improved to and $O(d)$-competitive 
algorithm for a tree of depth $d$ by Buchbinder {\em et al.}~\cite{BuchbinderFNT17}.
This problem generalizes the well-known TCP acknowledgment 
problem~\cite{KarlinKR03,DoolyGS01,BuchbinderJN07} 
and the online joint replenishment problem~\cite{BuchbinderKLMS13},
which represent special cases where the tree is
of depth $1$ and $2$ respectively. In the \osd problem, the server 
does not return to a specific location after each service, and can reside
anywhere on the metric space. Moreover, we consider arbitrary delay penalty 
functions, as against linear delays or deadlines.

\smallskip
\noindent {\bf Combining distances with delays.} 
Recently, Emek {\em et al.}~\cite{EmekKW16} suggested the problem 
of minimum cost online matching with delay. In this problem,
points arrive on a metric space over time and have to be matched
to previously unmatched points. The goal is to minimize the sum 
of the cost of the matching (length of the matching edges)
and the delay cost (sum of delays in matching the points).
%
%
Given an $n$-point metric space with aspect ratio
$\Delta$ (the ratio of the maximum distance to the minimum
distance), they gave a randomized algorithm with competitive ratio
of $O(\log^2n+\log\Delta)$, 
which was later improved to $O(\log  n)$ for general metric spaces \cite{AzarCK17}. 
The \osd problem has the same notion of combining distances and delays,
although there is no technical connection since theirs is a matching 
problem.




\section{Preemptive Service Algorithm}
\label{sec:single-server-algo}

In this section, we give an $O(h^3)$-competitive algorithm for the \osd problem on an HST
of depth $h$ (Theorem~\ref{thm:osd-hst}). We have already discussed how this extends to general metric 
spaces (Theorem~\ref{thm:osd}) via metric embedding. We call this the preemptive service 
algorithm, or \ps algorithm. The algorithm is described in detail in this section, and 
a summary of the steps executed by the algorithm is given in Fig.~\ref{alg:waiting}, \ref{alg:serving}, and \ref{alg:tf}.
The analysis of its competitive ratio appears in the next section. We will assume throughout,
wlog,\footnote{wlog = without loss of generality} that all requests arrive at leaf nodes
of the HST.

Note that an algorithm for the \osd problem needs to make two kinds of decisions: when to move the server, and what requests to serve when it decides to move the server. The \ps algorithm operates in two alternating modes that we call the {\em waiting phase} and the {\em serving phase}. The waiting phase is the default state where the server does not move and the delay of unserved requests increases with time. As soon as the algorithm decides to move the server, it enters the serving phase, where the algorithm serves a subset of requests by moving on the metric space and ends at a particular location. The serving phase is instantaneous, at the end of which the algorithm enters the next waiting phase. 

\begin{figure}
\begin{algorithmic}[1]
	\WHILE{no major edge is saturated}
    	\FOR{each unserved request $r$}
        	\STATE{Let $p_r$ be path connecting $r$ and the server. 
            Increment the counter on the unsaturated edge closest 
            to $r$ on path $p_r$ by $r$'s instantaneous delay penalty.}
        \ENDFOR
    \ENDWHILE
\end{algorithmic}
\caption{The waiting phase of the \ps algorithm}
\label{alg:waiting}
\end{figure}

\begin{figure}
\begin{algorithmic}[1]
	\STATE{Identify the set of key edges $K$.}
    \STATE{Initialize the set of edges to be traversed $S = \{e:e$ is in between the major edge and a key edge$\}$.}
    \FOR{each edge $e\in K$}
    	\STATE{Update $S = S \cup {\bf Time-Forwarding}(e)$.}
    \ENDFOR
	\STATE{Mark all requests reached by edges in $S$ as served.}
    \STATE{Let $e = (u, v)$ be the last key edge in DFS order on $S$. 
    Further, let the server's current location w.r.t. $e$ be on the side of $u$. 
    Then, update the server's location to $v$.}
\end{algorithmic}
\caption{The serving phase of the \ps algorithm}
\label{alg:serving}
\end{figure}

\begin{figure}
\begin{algorithmic}[1]
	\STATE{Initialize the output set of edges $S = \emptyset$.}
	\STATE{Let time $t$ be the first (future) time when $e$ is over-saturated
		by the unserved requests in $X_e$.}
    \FOR{each unserved request $r$ in $X_e$ that is critical at time $t$}
    	\STATE{Let $p_r$ be the path connecting $r$ to $e$.}
    	\STATE{Update $S = S \cup \{e'\in p_r\}$.}
    \ENDFOR	
    \IF{there exists some unserved request in $X_e$ that is not critical at time $t$}
		\STATE{Let $G(e)$ be the set of saturated edges that are not over-saturated 
        themselves, but their path to $e$ contains only over-saturated edges.}
		\STATE{Select $H(e) \subseteq G(e)$ such that $\sum_{e'\in H(e)} \ell(e') = \ell(e)$.} 
        \FOR{each edge $e'\in H(e)$}
			\STATE{Update $S = S \cup {\bf Time-Forwarding}(e')$.}
        \ENDFOR
   \ENDIF
   \RETURN{$S$.}
\end{algorithmic}
\caption{The subroutine {\bf Time-Forwarding}$(e)$ used in the \ps algorithm}
\label{alg:tf}
\end{figure}

\subsection{Waiting Phase} 
The \ps algorithm maintains a counter on each edge. The value of the counter on an edge $e$ at any time is called its {\em weight} (denoted $w(e)$). $w(e)$ starts at 0 at the beginning of the algorithm, and can only increase in the waiting phase. (Counters can be reset in the serving phase, i.e., $w_e$ set to 0, which we describe later.) At any time, the algorithm maintains the invariant that $w_e \leq \ell(e)$, where $l(e)$ is the length of edge $e$.

\smallskip\noindent
{\em When a counter reaches its maximum value, i.e., $w(e) = \ell(e)$, 
we say that edge $e$ is {\bf saturated}.}

\smallskip\noindent
During the waiting phase, every unserved request increases the counter on some edge on the path between the request and the current location of the server. We will ensure that this is always feasible, i.e., all edges on a path connecting the server to an unserved request will not be saturated. In particular, every unserved request increments its nearest unsaturated counter on this path by its instantaneous delay penalty. Formally, let $c_r(\tau)$ denote the delay penalty of a request $r$ for a delay of $\tau$. Then, in an infinitesimal time interval $[t, t+\epsilon]$, every unserved request $r$ (say, released at time $t_r$)  contributes $c^{'}_{r}(t-t_r) = c_{r}(t-t_r+\epsilon) - c_{r}(t-t_r)$ weight to the nearest unsaturated edge on its path to the server.
%

Now, we need to decide when to end a waiting phase. One natural rule would be to end 
the waiting phase when all edges on a request's path to the server are saturated. 
However, using this rule introduces some technical difficulties. Instead, note that in an HST,
the length of the entire path is at most $O(1)$ times the maximum length of an edge 
on the path. This allows us to use the maximum-length edge as a proxy for the entire path.

\smallskip\noindent
{\em For each request, we define its {\bf major edge} to be the longest edge on the path from this request to the server. If there are two edges with the same length, we pick the edge closer to the request.}

%
%
%
%
%

%
%

\smallskip\noindent
A waiting phase ends as soon as {\em any} major edge becomes saturated. If multiple major edges are saturated at the same time, we break ties arbitrarily. Therefore, every serving phase is triggered
by a unique major edge. 


\subsection{Serving Phase} 
In the serving phase, the algorithm needs to decide which requests to serve. A natural choice
is to restrict the server to requests whose major edge triggered the serving phase. This is 
because it might be prohibitively expensive for the server to reach other requests in this 
serving phase. 

\smallskip\noindent
{\em The {\bf relevant subtree} of an edge $e$ (denoted $R_e$) 
is the subtree of vertices for which $e$ is the major edge.}

\smallskip\noindent
We first prove that
the relevant subtree $R_e$ is a subtree of the HST, where one
of the ends of $e$ is the root of $R_e$. 

\begin{figure}
\centering
\includegraphics[width=.75\columnwidth]{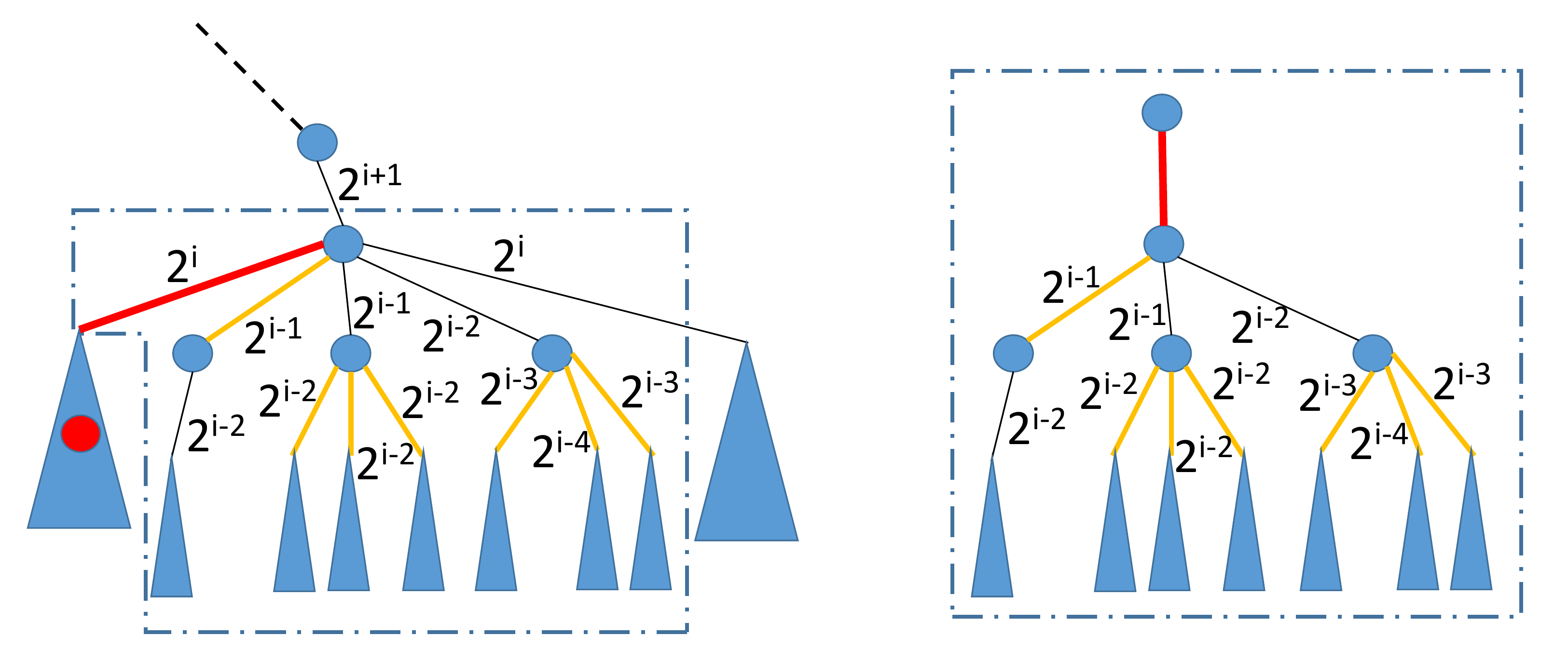}
\caption{\small Example where major edge is not the parent of the relevant subtree. 
The major edge is in red, the key edges in yellow, and the red circle is
the location of the algorithm's server.
In this situation, the server is in $T_e$ as shown and $X_e = R_e$. The 
dotted boxes show the edges in the critical subtree $C_e$. In particular,
the figure on the right shows the transformation of the critical subtree, 
where the major edge is now depicted as the parent of the relevant subtree.}
\label{fig:majoredge}
\end{figure}

\begin{lemma}\label{clm:majoredge}
The major edge is always connected to the root of the relevant subtree.
\end{lemma}
\begin{proof}
Let $r$ be the root of the relevant subtree (not including the major edge).
If the server is at a descendant of $r$, 
then by structure of the HST, the longest edge must be the first edge 
from $r$ to the server (this is the case in Fig.~\ref{fig:majoredge}). 
If the server is not at a descendant of $r$, then the longest edge 
from the relevant subtree to the server must be the edge connecting 
$r$ and its parent. In either case, the lemma holds.
\end{proof}

\begin{figure}
\centering
\includegraphics[width = \columnwidth]{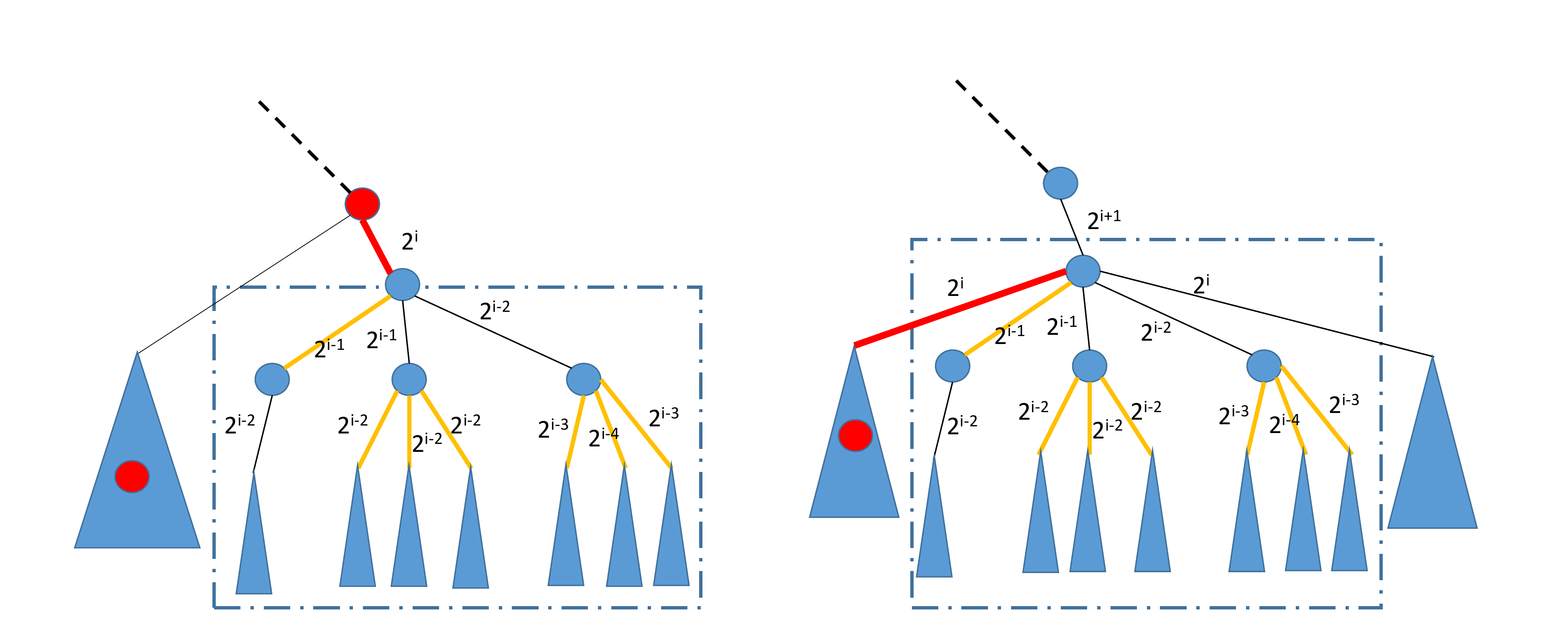}
\caption{\small The two possible configurations of the major edge (in red) vis-\`{a}-vis the 
relevant subtree. The subtree surrounded by dashed box is the relevant subtree. 
The red nodes are possible locations of the server. The edges in yellow are key edges.
In the figure to the right, note that the edge of length $2^i$ to the right is 
{\em not} included in the relevant subtree.}
\label{fig:keyedge}
\end{figure}

There are two possible configurations
of the major edge $e$ vis-\`{a}-vis its relevant subtree $R_e$. 
Either $e$ is the parent of the top layer of edges in $R_e$, 
or $e$ is a sibling of this top layer of edges (see Fig.~\ref{fig:keyedge} 
for examples of both situations). In either case, the definition of 
major edge implies that the length of $e$ is at least $2$ times the 
length of {\em any} edge in $R_e$.


The \ps algorithm only serves requests in the relevant subtree $R_e$
of the major edge $e$ that triggered the serving phase.
Recall from the introduction that the algorithm
performs two kinds of service: critical service and preemptive service. 
First, we partition unserved requests in $R_e$ into two groups. 

\smallskip\noindent
{\em Unserved requests in the relevant subtree $R_e$ that are connected to the 
major edge $e$ by saturated edges are said to be {\bf critical}, 
while other unserved requests in $R_e$ are said to be {\bf non-critical}.}

\smallskip\noindent
The algorithm serves {\em all} critical requests in $R_e$ in the serving phase. 
This is called critical service. 
In addition, it also serves a subset of non-critical requests in $R_e$ 
as preemptive service. Next, we describe the selection of non-critical 
requests for preemptive service.

\smallskip\noindent
{\bf Preemptive Service.}
We would like to piggyback the cost of preemptive service on critical service. 
Let us call the subtree connecting the major edge $e$ to the critical requests
in $R_e$ the {\em critical subtree} $C_e$ (we also include $e$ in $C_e$). 
Recall that $e$ may either be a sibling or the parent of the 
top layer of edges in $R_e$ (Fig.~\ref{fig:keyedge}). Even in the first case,
in $C_e$, we make $e$ the parent of the top layer of edges in $R_e$ 
for technical reasons
(see Fig.~\ref{fig:majoredge} for an example of this transformation). 
Note that this does not
violate the HST property of $C_e$ since the length of $e$ is
at least $2$ times that of any other edge in $R_e$, by virtue of $e$ being
the major edge for all requests in $R_e$.


The cost of critical service is the total length of edges in the critical
subtree $C_e$. However, using this as the cost bound
for preemptive service in the \ps algorithm 
causes some technical difficulties. Rather, we identify
a subset of edges (we call these {\em key edges}) in $C_e$
whose total length is at least $1/h$ times that of all edges in $C_e$. 
We define key edges next. 

Let us call two edges {\em related} if they form an 
ancestor-descendant pair in $C_e$ . A {\em cut} in $C_e$ is a
subset of edges, no two of which are related, and whose removal disconnects 
all the leaves of $C_e$ from its root.

\smallskip\noindent
{\em The set of {\bf key edges} is formed by the cut of maximum total edge length
in the critical subtree $C_e$. If there are multiple such cuts, any one of them is
chosen arbitrarily.}

\smallskip\noindent
The reader is referred to Fig.~\ref{fig:keyedge} for an example of key edges.
Note that either the major edge $e$ is the unique key edge, or the key edges
are a set of total length at least $\ell(e)$ whose removal disconnects $e$ 
from all critical requests in the relevant subtree $R_e$. It is
not difficult to see that the total length of the key edges is at least $1/h$ 
times that of all edges in $C_e$.

We will perform {\em recursive time forwarding} on the key edges to identify the 
requests for preemptive service. Consider any single key edge $e$. Let us 
denote the subtree below $e$ in the HST by $T_e$. Typically, when we are 
forwarding time on a key edge $e$, we consider requests in subtree $T_e$. 
However, there is one exception: if $e$ is the major edge and is a sibling 
of the top layer of its relevant subtree (Fig.~\ref{fig:keyedge}), then
we must consider requests in the relevant subtree $R_e$. We define a
single notation for both cases:

\smallskip\noindent
{\em For any key edge $e$, let $X_e$ denote the following:
\begin{itemize}\itemsep0pt
	\item If the algorithm's server is not in $T_e$, then $X_e = T_e$.
	\item If the algorithm's server is in $T_e$, then $X_e = R_e$.
\end{itemize}}

\smallskip\noindent
To serve critical requests, the server must traverse a key edge $e$ 
and enter $X_e$. Now, suppose $X_e$ looks exactly 
like Fig.~\ref{fig:preempt}, i.e., there is a single critical request and
many non-critical requests. As we saw in the introduction, the server cannot
afford to only serve the critical request, but must also do preemptive service.
Our basic principle in doing preemptive service is to simulate future time
and serve requests in $X_e$ that would become critical the earliest. 
So, we forward time into the future, which causes the critical subtree in
$X_e$ to grow. But, our total budget for preemptive service
in $X_e$ is only $\ell(e)$, the length of $e$. (This is because the overall 
budget for preemptive service in this serving phase 
is the sum of lengths of all the key edges.
This budget is distributed by giving a budget equal to its length to each 
key edge.) This budget cap implies that we should stop time
forwarding once the critical subtree in $X_e$ has total edge length $\ell(e)$;
otherwise, we will not have enough budget to serve the requests that 
have become critical after time forwarding.
As was the case earlier, working with the entire critical 
subtree in $X_e$ is technically difficult. Instead, we forward time until
there is a cut in the critical subtree of $X_e$ whose total edge length is equal to $\ell(e)$ 
(at this point, we say edge $e$ is over-saturated).  
This cut now forms the next layer of edges in the recursion (we
call these {\em service edges}). Note that the key edges form the first 
layer of service edges in the recursion. Extending notation,
$X_{e'} = T_{e'}$ for all service edges in subsequent layers of recursion.
We recursively forward time on each of these service edges $e'$ in their 
respective subtrees $X_{e'}$ using the same process. 

How does the recursion stop? Note that if the current service edge $e'$
is a leaf edge, then the recursion needs to stop since the only 
vertex in $X_{e'}$ is already critical at the time the previous
recursive level forwarded to. More generally, if all requests in 
$X_{e'}$ become critical without any cut in the critical subtree of 
$X_{e'}$ attaining total length $\ell(e')$, 
then we are assured that the cost of serving {\em all} requests in $X_{e'}$
is at most $h$ times the cost of traversing edge $e'$. 
In this case, we do not need to forward 
time further on $e'$ (indeed, time forwarding does not 
change the critical subtree since all requests are already critical), 
and the recursion ends. 

We now give a formal description of this recursive time forwarding process. 
For any edge $e$, let $\child(e)$ be the edges in the first layer of 
subtree $X_e$. (When $X_e = T_e$, these are simply the children edges
of $e$. But, in the special case where $X_e = R_e$, these are the 
siblings of $e$ of smaller length than $e$.) We define
\begin{equation*}
	f(e) = \begin{cases}
				\max\left(\ell(e), \sum_{e_i \in \child(e)}{f(e_i)}\right) & \text{if } e \text{ is saturated.}\\
                0 & \text{otherwise.}
            \end{cases}
\end{equation*}        
This allows us to formally define over-saturation, that we intuitively described
above.

\smallskip\noindent
{\em An edge $e$ is said to be {\bf over-saturated} if either of the following
happens:
\begin{itemize}\itemsep0pt
\item $e$ is saturated and $\sum_{e_i \in \child(e)}{f(e_i)} \geq \ell(e)$, 
i.e., $f(e)$ is determined by $\sum_{e_i \in \child(e)}{f(e_i)}$ in
the formula above  (we call this over-saturation by children).
\item All requests in $X_e$ are connected to $e$ via saturated edges.
\end{itemize}
}

\smallskip\noindent
Using this notion of over-saturation, the \ps algorithm now identifies 
a set of edges $S$ that the server will traverse in the relevant subtree
$R_e$ of the major edge $e$ that triggered the serving phase. 
In the formal description,
we will not consider critical service separately since the edges in 
the  critical subtree $C_e$ will be the first ones to be added to
$S$ in our recursive {\bf Time-Forwarding}
process. Initially, all saturated edges between the key edges 
and the major edge, both inclusive, are added to $S$. This ensures 
that the algorithm traverses each key edge. Now, the 
algorithm performs the following recursive process on every 
key edge $e'$ in subtree $X_e$.

Intuitively, the edges in $G(e)$ are the ones
that form the maximum length cut in the critical subtree of $X_e$
at time $t$ (we show this property formally in Lemma~\ref{lma:tf}).
In the recursive step, ideally we would like to recurse on all edges in $G(e)$. 
However, if $e_l$ is the last edge added to $G(e)$ during {\bf Time-Forwarding}($e$), 
it is possible that $\sum_{e'\in G(e)\setminus \{e_l\}} \ell(e') < f(e)$ but 
$\sum_{e'\in G(e)} \ell(e') > f(e)$. In this case, we cannot afford to recurse
on all edges in $G(e)$ since their total length is greater than that of $e$.
This is why we select the subset $H(e)$ whose sum of edge lengths is exactly equal 
to that of $e$. We show in Lemma~\ref{lma:subset} below that this is always possible to do.
The set of edges that we call {\bf Time-Forwarding} on, i.e., the key edges and
the union of all $H(e)$'s in the recursive algorithm, are called {\bf service edges}.

\begin{lemma} \label{lma:subset}
Let $e$ be an edge oversaturated by children. 
$G(e)$ represents a set of saturated but not over-saturated edges
satisfying: 
$$\sum_{e' \in G(e)} \ell(e') = \sum_{e' \in G(e)} f(e') \geq \ell(e).$$  
Then, there is some subset $H(e) \subseteq G(e)$ 
such that $\sum_{e'\in H(e)} \ell(e') = \ell(e)$.
\end{lemma}
\begin{proof}
Note that all edge lengths are of the form $2^i$ for non-negative integers 
$i$. Furthermore, the length of every edge in $G(e)$ is strictly smaller
than the length of $e$. Thus, the lemma is equivalent to showing that for 
any set of elements of the form $2^i$ (for some $i \leq j-1$) that sum to at least
$2^j$, there exists a subset of these elements that sum to exactly $2^j$. 

First, note that we can assume that the sum of these elements is strictly 
smaller than $(3/2)\cdot 2^j$. If not, we repeatedly discard an arbitrarily
selected element from the set until this condition is met.
Since the discarded value is at most $2^{j-1}$ at each stage, the sum 
cannot decrease from $\geq (3/2)\cdot 2^j$ to $< 2^j$ in one step.

Now, we use induction on the value of $j$. For $j=1$, the set can only 
contain elements of value $2^0 = 1$, and sums to the range $[1,3)$. Clearly, 
the only option is that there are two elements, both of value $1$. 
For the inductive step, order the elements in decreasing value.
There are three cases:
\begin{itemize}
\item  If there are $2$ elements of value $2^{j-1}$, we output
these $2$ elements.
\item If there is $1$ element of value $2^{j-1}$, we add it to
our output set. Then, we start discarding elements in arbitrary order 
until the remaining set has total value less than 
$(3/2)\cdot 2^{j-1}$. Then, we apply the inductive hypothesis on
the remaining elements. 
\item Finally, suppose there is no element of value $2^{j-1}$ in
the set. We greedily create a subset by adding elements to the 
subset in arbitrary order until their sum is at least $2^{j-1}$.
Since each element is of value at most $2^{j-2}$, the total
value of this subset is less than $(3/2)\cdot 2^{j-1}$. Call this 
set A. Now, we are left with elements summing to total value at 
least $2^{j-1}$ and less than $2^j$. We again repeat this process
to obtain another set $B$ of total value in the range 
$[2^{j-1}, (3/2)\cdot 2^{j-1})$. We now recurse on these two 
subsets $A$ and $B$.\qedhere
\end{itemize}
\end{proof}

Once all chains of recursion have ended, the server does a DFS on the edges in $S$, 
serving all requests it encounters in doing so. 
After this, the server stops at the bottom of the 
last key edge visited.

\smallskip
\noindent
{\em Remark:} If the HST is an embedding of a general metric space, 
then the leaves in the HST are the only actual locations on the metric space. So,
when the \ps algorithm places the server at an intermediate vertex, the server
is actually sent to a leaf in the subtree under that vertex in the actual metric
space. It is easy to see
that this only causes a constant overhead in the service cost. Note that this 
does not affect the \ps algorithm's operation at all; it operates as if the server
were at the non-leaf vertex.


\smallskip
Finally, the algorithm resets the counters, i.e., sets $w(e) = 0$, 
for all edges $e$ that it traverses in the serving phase.

This completes the description of the algorithm. 


%
%
%
%
%
%
%

\section{Competitive Ratio of Preemptive Service Algorithm}
\label{sec:single-server-analysis}

In this section, we show that the \ps algorithm described in the previous section 
has a competitive ratio of $O(h^3)$ on an HST of depth $h$ (Theorem~\ref{thm:osd-hst}), 
which implies a competitive ratio of $O(\log^4 n)$ on general metric spaces 
(Theorem~\ref{thm:osd}).


First, we specify the location of requests that can put weight on an edge
at any time in the \ps algorithm. 

\begin{lemma}
\label{lma:history}
	At any time, for any edge $e$, the weight on edge $e$
		is only due to unserved requests in $X_e$.
\end{lemma}
\begin{proof}
	When the server serves a request, it traverses all edges on
    its path to the request and resets their respective counters.
    So, the only possibility of violating the lemma is due to a change
    in the definition of $X_e$. But, for the definition of $X_e$ to
    change, the server must traverse edge $e$ and reset its counter.
    This completes the proof.
\end{proof}

Next, we observe that the total delay penalty of the algorithm is bounded by its 
service cost. (Recall that the service cost is the distance moved by the server.)

\begin{lemma} \label{waitingcosts}

The total delay penalty is bounded by the total service cost.

\end{lemma}

\begin{proof}
By design of the algorithm, every unserved request always increases the counter 
on some edge with its instantaneous delay penalty. The lemma now follows from
the observation that the counter on an edge $e$ can be reset from 
at most $\ell(e)$ to $0$ only when
the server moves on the edge incurring service cost $\ell(e)$.
\end{proof}


Next, we show that the total service cost can be bounded against the 
length of the key edges.
First, we formally establish some properties of the {\bf Time-Forwarding}
process that we intuitively stated earlier.

\begin{lemma}
\label{lma:tf}
	When the algorithm performs {\bf Time-Forwarding}($e$), then the 
    following hold:
    \begin{itemize}\itemsep0pt
    	\item If a saturated edge $e'\in X_e$ is not over-saturated 
        by its children, all cuts in the saturated subtree in $X_{e'}$ 
        have total edge length at most $\ell(e')$.
    	\item $\sum_{e'\in G(e)} \ell(e') \leq (3/2) \ell(e)$.
        \item $G(e)$ is the cut of maximum total edge length in 
        the saturated subtree of $X_e$ at time $t$, the time when $e$ is first over-saturated. 
    \end{itemize}
\end{lemma}
\begin{proof}
	The first property follows from the fact that for a saturated
    edge $e'$, the definition of the function $f(.)$ implies that 
    the value of $f(e')$ is at least the sum of lengths of any cut in 
    the saturated subtree of $X_{e'}$ at time $t$. 
   
	For the second property, note that every edge in $X_e$ has 
    length at most $\ell(e)/2$, and use Lemma~\ref{lma:subset}.
    
    Now, suppose the third property is violated, and there is a larger 
    saturated cut $G'(e)$. Since $G(e)$ over-saturates all its ancestor 
    edges, it follows that for any such ancestor edge $\tilde{e}$
    in $X_e$, the descendants of $\tilde{e}$
    in $G(e)$ have total length at least $\ell(\tilde{e})$.
    Thus, there must 
    be an edge $e^*\in G(e)$ such that the sum of lengths of the 
    descendants of $e^*$ in $G'(e)$ is greater than $\ell(e^*)$.
    But, if this were the case, $e^*$ would be over-saturated,
    which contradicts the definition of $G(e)$.
\end{proof}

\begin{lemma} \label{keyedge}

The total service cost is at most $O(h^2)$ times the sum of lengths of key edges.

\end{lemma}

\begin{proof}
By the HST property, the distance traveled
by the server from its initial location to the major edge $e$ is at most $O(1)$ times
$\ell(e)$. But, $\ell(e)$ is at most the total length of the key edges, 
since $e$ itself is a valid cut in the critical subtree $C_e$. 
Hence, we only consider server movement in the relevant subtree $R_e$. 

Now, consider the edges connecting the major edge to the key edges.
Let the key edges have total length $L$. Consider the cut comprising the 
parents of all the key edges. They also form a valid cut, and hence, their
total length is at most $L$. We repeat this argument until we get to the 
major edge. Then, we have at most $h$ cuts, each of total edge length at most 
$L$. Thus the total length of all edges connecting the major edge
to the key edges is at most $h$ times the length of the key edges.

%
%
%
%
%

This brings us to the edges traversed by the server in the subtrees $X_e$, 
for key edges $e$. We want to bound their total length by $O(h^2\cdot \ell(e))$.
We will show a more general property: 
that for {\em any service edge} $e$, the total length
of edges traversed in $X_e$ is $O(h^2\cdot \ell(e))$. First, we consider the case that
the recursion ends at $e$, i.e., $e$ was not over-saturated by its children.
By Lemma~\ref{lma:tf}, this implies that every cut in the saturated subtree
in $X_e$ has total length at most $\ell(e)$. In particular, we can apply 
this to the layered cuts (children of the key edges, their children,
and so on) in the saturated subtree. There are at most
$h$ such cuts covering the entire saturated subtree, implying that the total
cost of traversing the saturated edges in $X_e$ is $O(h\cdot \ell(e))$.

Next, we consider the case when $e$ is over-saturated by its children.
We use an inductive argument on the height of the subtree $X_e$. 
Suppose this height is $h$. Further, let
the edges in $H(e)$ (that we recurse on in {\bf Time-Forwarding}($e$)) 
be $e_1, e_2, \ldots$, where 
the height of subtree $X_{e_i}$ is $h_i \leq h-1$. 
By definition of $H(e)$, we have $\sum_i \ell(e_i) = \ell(e)$. 
Using the inductive hypothesis, the total 
cost inside subtree $X_{e_i}$ is at most $C\cdot h_i^2 \cdot \ell(e_i)$
for some constant $C$.
Summing over $i$, and using $h_i \leq h-1$, 
we get that the total cost is at most $C\cdot (h-1)^2 \cdot \ell(e)$.
This leaves us with the cost of reaching the critical requests
in $X(e')$ for each $e'\in G(e)$, from edge $e$. We can partition these edges into at most
$h$ layers, where by Lemma~\ref{lma:tf}, the total cost of 
each such layer is at most $(3/2) \cdot \ell(e)$. 
It follows that the total cost of all layers
is at most $(3/2)\cdot h \cdot \ell(e)$. Adding this to the 
inductive cost, the total cost is at most $C\cdot h^2 \cdot \ell(e)$
for a large enough $C$.
This completes the inductive proof.
\end{proof}


Now, we charge the lengths of all key edges to the optimal solution \opt.
Most of the key edges will be charged directly to \opt, but we need
to define 
%
%
%
a potential function $\Phi$ for one special case. For a fixed optimal solution, we define $\Phi$ as
$$
\Phi = \kappa d(\mbox{alg's server}, \mbox{\opt's server}).
$$
In words, $\Phi$ is equal to $\kappa$ 
times the distance between algorithm's server and \opt's server. Here, $\kappa = ch^2$ for a large enough constant $c$. 
%
Overall, we use a charging argument to bound all costs and potential changes in our algorithm to 
\opt such that the charging ratio is $O(h^3)$.

The first case is when the major edge $e$ is the only key edge, 
and \opt's server is in $X_e$. 
Then, all costs are balanced by the decrease in potential. 

\begin{lemma}\label{clm:potentialdecrease}
When the major edge $e$ is the only key edge, and \opt's server is in $X_e$,
the sum of change in potential function $\Phi$ and the service cost is non-positive.
\end{lemma}
\begin{proof}
This is immediate from the definition of $\Phi$ and Lemma~\ref{keyedge}. 
\end{proof}

In other cases, we charge all costs and potential changes 
to either \opt's waiting cost or \opt's moving cost 
by using what we call a {\em charging tree}. 

\smallskip\noindent
{\em 
A \textbf{charging tree} is a tree of (service edge, time interval) pairs. 
For each service edge $e$ at time $t$, let $t'$ be the last time that 
$e$ was a service edge ($t' = 0$ if $e$ was never a service edge in the past). 
Then, $(e, [t', t))$ is a node in a charging tree.
We simultaneously define a set of charging trees, the roots of which exactly
correspond to nodes $(e, [t', t))$ where $e$ is a key edge at time $t$.

To define the charging trees, we will specify the children of a node
$(e, [t', t))$:
\begin{itemize}\itemsep0pt
\item The node $(e, [t',t))$ is a leaf in the charging tree 
if $t' = 0$ or $e$ is not over-saturated by children at time 
$t'$ (i.e., all previously unserved requests in $X_e$ were served at time $t'$).
\item Otherwise, let $H(e)$ be as defined in {\bf Time-Forwarding}($e$). 
For any $e' \in H(e)$, let $t''$ be the last time that $e'$ was a service edge. Then, 
for each $e'\in H(e)$, $(e',[t'',t'))$ is a child of $(e,[t',t))$.
\end{itemize}
}
%
%
%
%

\begin{figure}
\centering
\includegraphics[width=0.55\columnwidth]{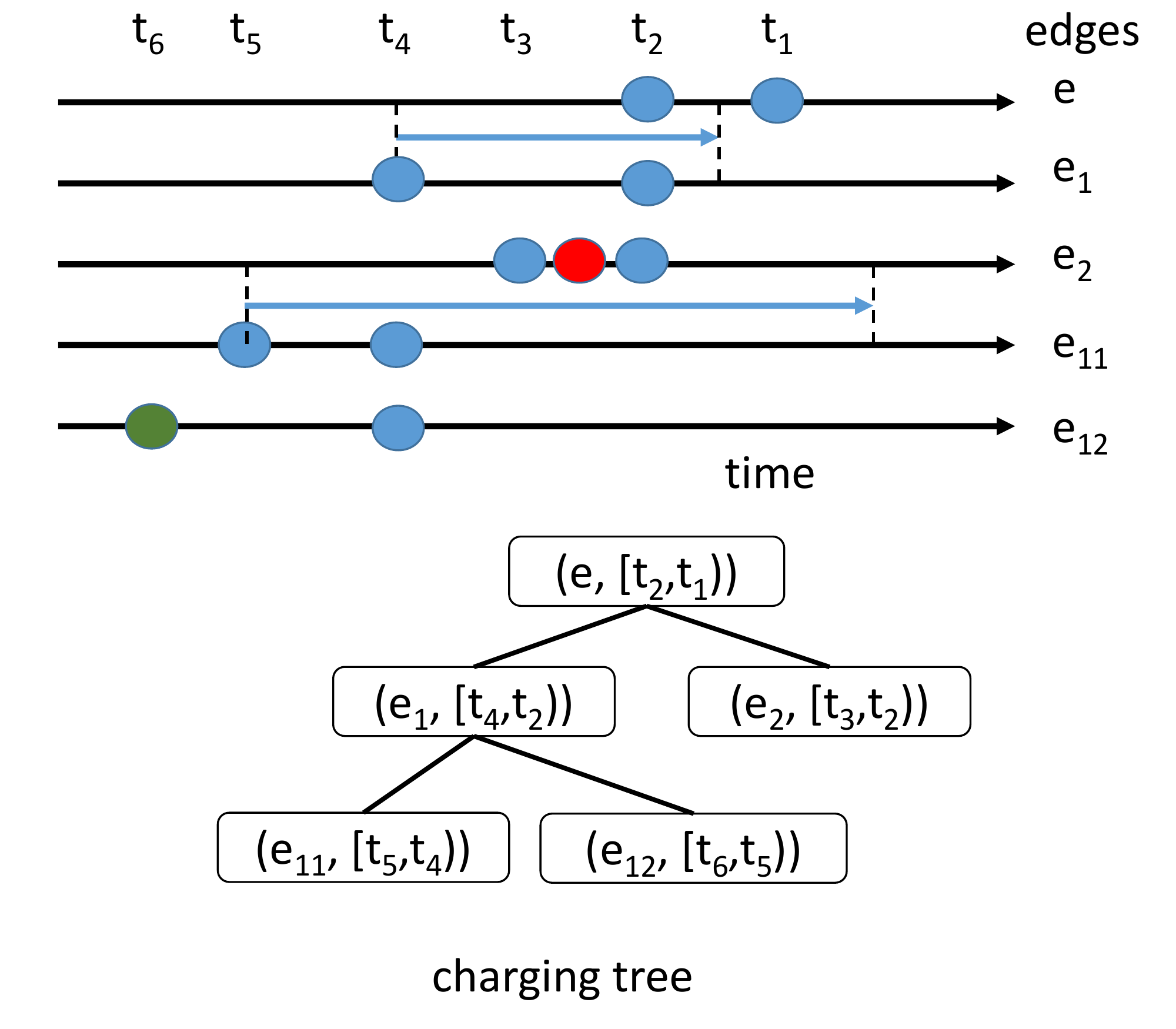}
\caption{\small Example of a charging tree. 
On the top is the timeline corresponding to the charging tree on the bottom.
The blue circles represent that the corresponding edge is a service edge in 
the serving phase of the algorithm at that time. The red circle indicates that \opt's server traverses the edge in 
the interval defined the two blue circles surrounding the red circle. The green circle indicates that the corresponding
edge is not only a service edge in the serving phase of the algorithm at that time, but that the algorithm actually
served all requests in the corresponding subtree $X_{e_{12}}$.}\label{fig:chargingtree}
\end{figure}


\smallskip\noindent
See Fig.~\ref{fig:chargingtree} for an example. This figure shows a 
charging tree with 5 nodes: $(e,[t_2,t_1))$, $(e_1,[t_4,t_2))$, 
$(e_2,[t_3,t_2))$, $(e_{11},[t_5,t_4))$, $(e_{12},[t_6,t_4))$. 

We say that \opt incurs {\em moving cost} at a charging tree node $(e,[t',t))$ 
if it traverses edge $e$ in the time interval $[t', t)$. 
Next, we incorporate \opt's delay penalties in the charging tree. 
%
%
We say that \opt incurs {\em waiting cost} of $\ell(e)$ at a 
charging tree node $(e, [t', t))$, 
if there exists some (future) time $t^* \geq t$ such that:
\begin{itemize}\itemsep0pt
	\item \opt's server is never in $X_e$ in the time interval $[t', t^*)$, and
	\item the requests that arrive in $X_e$ in the time interval $[t', t)$ 
	have a total delay penalty of at least $\ell(e)$ if they remain unserved at time $t^*$. 
\end{itemize}
Intuitively, if these conditions are met, then \opt has a delay penalty of 
at least $\ell(e)$ from only the requests in subtree $T_e$ that arrived in 
the time interval $[t', t)$.


Without loss of generality, we assume that at time 0, \opt traverses 
all the edges along the paths between the starting location of the 
server and all requests.
 (Since these edges need to be traversed at least once by {\em any} solution, 
this only incurs an additive cost in \opt, 
and does not change the competitive ratio by more than a constant factor.)

Now, we prove the main property of charging trees.

\begin{lemma}[Main Lemma] \label{lem:chargingtree}
For any key edge $e$ served at time $t$ by the algorithm, if \opt's server 
was not in $X_e$ at time $t$, then \opt's waiting and moving cost in the charging tree rooted at 
$(e,[t',t))$ is at least $\ell(e)$.
\end{lemma}

\begin{figure}
\centering
\includegraphics[width=\columnwidth]{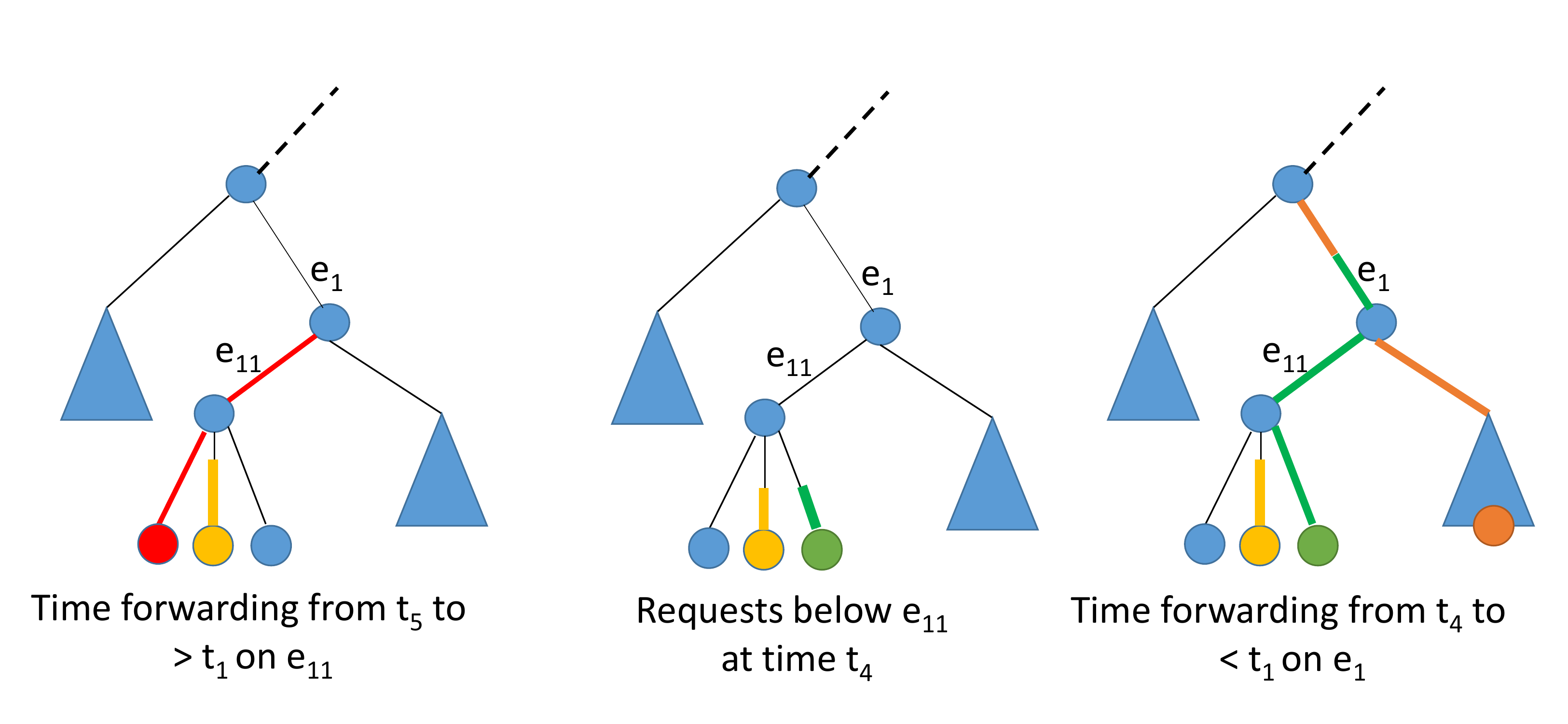}
\caption{\small Illustration of {\bf Case 2} in Lemma~\ref{lem:chargingtree}. The left figure shows the time forwarding for $e_{11}$ at time $t_5$. All the red requests are served at $t_5$. The center figure shows the waiting time below $e_{11}$ at time $t_4$. Note that since $t_4 < t$, the yellow requests could not reach $e_{11}$, and the green requests are requests that came between $t_5$ and $t_4$. The right figure shows the time forwarding process for $e_1$ at time $t_4$. We know $e_{11}$ must be saturated because it is a child in charging tree. However, the yellow requests cannot contribute to $e_{11}$ because they do not even reach $e_{11}$ at time $t$. Therefore, all the weight on $e_{11}$ must come from green requests.}\label{fig:timeforward}
\end{figure}

\begin{proof} 
We use the recursive structure of the charging tree. Suppose the root of the tree is $(e,[t',t))$, 
 the current node we are considering is $(u, [t_u',t_u))$, and the parent of the current 
node is $(v,[t_v', t_v))$. (Here, $u$ and $v$ are edges in the HST.) By the structure of the charging tree, we know that $t_v'=t_u$. We will 
now list a set of conditions. At node $(u, [t_u',t_u))$, if one of these conditions holds, then 
we show that the waiting or moving cost incurred by \opt at this node is at least $\ell(u)$,
the length of edge $u$. Otherwise, we recurse on all children of this charging tree node. The 
recursion will end with one these conditions being met.

\smallskip\noindent
{\bf Case 1.}
{\em \opt crossed edge $u$ in the time interval $[t_u',t_u)$:} 
This is the case for $(e_2,[t_3,t_2))$ in Fig.~\ref{fig:chargingtree}. 
In this case, \opt incurs moving cost of $\ell(u)$ at node $(u, [t_u', t_u))$
of the charging tree.

\smallskip\noindent
{\bf Case 2.}
{\em Case 1 does not apply, and the algorithm forwarded time to greater than 
$t$ in {\bf Time-Forwarding}($u$) at time $t_u'$:}
This is the case for $(e_{11},[t_5,t_4))$ in Fig.~\ref{fig:chargingtree}.
A detailed illustration of this case is given in Fig.~\ref{fig:timeforward}.
Note that Case 1 does not apply to $(u, [t_u', t_u))$ by definition. 
Moreover, Case 1 also does not apply to any of the ancestors of this node in the charging tree; 
otherwise, the recursive charging scheme would not have considered the $(u, [t_u', t_u))$ node. 
Furthermore, at time $t$, \opt's server was not in $X_e$ by the condition of the 
lemma. These facts, put together, imply that \opt could not have served any request in 
$X_u$ in the time interval $[t_u', t)$ . In particular, this implies that \opt could not have 
served requests in $X_u$ that arrived in the time interval $[t_u',t_u)$ before time $t$(these are green requests in Fig.~\ref{fig:timeforward}).

We now need to show that these requests have a total delay penalty of at least $\ell(u)$
by time $t$. Since {\bf Time-Forwarding}($u$) forwarded time beyond $t$ at time $t_u'$, 
any request in $X_u$ that arrived before $t_u'$ and put weight on edge $u$ (these are red requests in Fig.~\ref{fig:timeforward}) must have 
been served at time $t_u'$. The requests left unserved (yellow requests in Fig.~\ref{fig:timeforward}) do not put weight on edge $u$ 
even after forwarding time to $t$.

Now, note that no ancestor of the $(u, [t_u', t_u))$ node in the charging tree 
can be in Case 2, else the recursive charging scheme would not consider the $(u, [t_u', t_u))$ 
node. Therefore, {\bf Time-Forwarding}($v$) must have forwarded time to at most $t$ 
at time $t_v' = t_u$. 
Suppose we reorder requests so that the requests that arrived earlier
put weight first (closer to the request location), 
followed by requests that arrived later. Then, the requests in $X_u$
that arrived before $t_u'$ but were not served at time $t_u'$ (yellow requests in Fig.~\ref{fig:timeforward})
do not put weight on edge $u$ in {\bf Time-Forwarding}($v$) at time 
$t_v'$, since time is being forwarded 
to less than $t$. Therefore, the only requests in $X_u$ that put weight
on edge $u$ in {\bf Time-Forwarding}($v$) at time $t_v'$ arrived in the time interval
$[t_u', t_u)$ (green requests in Fig.~\ref{fig:timeforward}). By Lemma~\ref{lma:history}, at any time,
the only requests responsible for the weight on edge $u$ are requests in $X_u$. Therefore, 
the requests that arrived in $X_u$ in the time interval $[t_u', t_u)$ saturate edge 
$e$ in {\bf Time-Forwarding}($v$) at time $t_v'$. 
In other words, these requests incur delay penalties 
of at least $\ell(u)$ if they are not served till time $t$. 
Since \opt did not serve any of these requests 
before time $t$, it follows that \opt incurs waiting cost of $\ell(u)$ at node
$(u, [t)u', t_u))$ of the charging tree.



\smallskip\noindent
{\bf Case 3.}
{\em Case 1 does not apply, and all requests in $X_u$ 
were served by the algorithm at time $t_u'$:}
This is the case for $(e_{12},[t_6,t_4))$ in Fig.~\ref{fig:chargingtree}. 
Similar to the previous case, {\bf Time-Forwarding}($v$) at time $t_v'$ 
must have been to at most time $t$. Since all requests in $X_u$ that 
arrived before time $t_u'$ were served at time $t_u'$, the requests 
that saturated edge $u$ in {\bf Time-Forwarding}($v$) at time $t_v' = t_u$ 
must have arrived in the time interval $[t_u',t_u)$ (same argument as
in Case 2). As in Case 2, 
\opt could not have served these requests before time $t$, and hence 
incurs delay penalty of at least $\ell(u)$ on these requests. Therefore,
\opt incurs waiting cost of $\ell(u)$ at node $(u, [t_u', t_u))$ of the 
charging tree.

By the construction of charging tree, the leaves are either in Case 3, 
or has starting time $0$ (which is in Case 1). So, this recursive argument 
always terminates.

Note that each key edge is the root of a charging tree, and for each node 
of the charging tree, the length of the corresponding edge is equal to 
the sum of lengths of the edges corresponding to its children
nodes. Therefore when the argument terminates, we can charge the cost 
of the key edge to a set of nodes in the charging tree whose total cost 
is exactly equal to the cost of the key edge. 
\end{proof}


So, we have shown that the length of key edges can be bounded against
\opt's cost on the charging tree. But, it is possible that in this 
analysis, we are using the same cost incurred by \opt in multiple
charging tree nodes. In other words, we need to bound the cost incurred
by \opt at charging tree nodes against the actual cost of \opt.

\begin{lemma}
\label{lma:opt-chargingtree}
	The total number of charging tree nodes on which the same request
	causes \opt to incur cost is at most $h+1$.
\end{lemma}
\begin{proof}
First, note that the charging tree nodes corresponding to the same 
edge $e$ represent disjoint time intervals. This implies that for
a given request that arrived at time $t$, \opt only incurs cost in 
the charging tree node $(e, [t_1, t_2))$ where $t\in [t_1, t_2)$. 
We now show that the total number of different
edges for which the same request is counted toward \opt's cost 
at charging tree nodes is at most $h+1$.

To obtain this bound, we observe that there are two kinds of 
edges we need to count: (1) edges on the path from the request to
the root of the HST, and (2) edges for which the request is in 
the relevant subtree. Clearly, there
are at most $h$ category (1) edges. But, there 
can be a much larger number of category (2) edges. 
However, we claim that
among these category (2) edges, \opt incurs cost in 
the charging tree at a node corresponding to at most 
one edge. To see this, we observe
the following properties:
\begin{itemize}\itemsep0pt
	\item If \opt incurs cost at a charging tree node
	corresponding to a category (2) edge $e$, then \opt's server
	must be in $T_e$ at the arrival time of the request.
	\item The category (2) edges are unrelated in the HST, 
	i.e., for any two such edges $e_1, e_2$, we have 
    the property that $T_{e_1}$ and $T_{e_2}$ are disjoint.\qedhere
\end{itemize}
\end{proof}

Finally, we are ready to prove the main theorem (Theorem~\ref{thm:osd-hst}).

\begin{proof}[Proof of Theorem~\ref{thm:osd-hst}]
Recall that we bound both the delay penalties and service cost of the algorithm by the total length of key edges times $O(h^2)$, in Lemmas~\ref{waitingcosts} and \ref{keyedge}. Now, we will charge the total length of key edges to $O(h)$ times \opt's costs. Consider a serving phase at time $t$. 

\smallskip\noindent
{\bf Case (a).}
Suppose \opt's server is not in subtree $X_e$ of a key edge $e$. Then, by 
Lemma~\ref{lem:chargingtree}, \opt incurs a cost of at least $\ell(e)$ in the charging tree rooted at $(e, [t', t))$, where $t'$ is the last time when $e$ was a service edge. On the other hand,  \opt's moving cost can be counted by at most one node in all charging trees, and by Lemma~\ref{lma:opt-chargingtree}, \opt's waiting cost can be counted by at most $h+1$ nodes in all charging trees. Therefore the total length of {\bf Case (a)} key edges is bounded by $(h+1)$ times the cost of \opt.

\smallskip\noindent
{\bf Case (b).}
Now, let us consider the situation where \opt's server is in the subtree $X_e$ for a key edge $e$. There are two possibilities:
\begin{itemize}
\item
{\bf Case (b)(i).}
If $e$ is also the major edge, it must be the unique key edge and we apply Lemma~\ref{clm:potentialdecrease}. 
\item
{\bf Case (b)(ii).}
Otherwise, $e$ is not the major edge. Then, any single key edge can be of length at most half the major edge, but the key edges add up to at least the length of the major edge. Thus, any single key edge only accounts for at most half the total length of key edges. Now, \opt's server can be in the subtree $T_e$ for only one of the key edges $e$ (note that in this case, $T_e = X_e$ since the major edge is not the unique key edge). It follows that the key edges in {\bf Case (a)} account for at least half the total length of the key edges. In this case, we can charge the {\bf Case (b)} key edges to the {\bf Case (a)} key edges (which are, in turn, charged to \opt by the argument above).
\end{itemize}

Finally, we also need to bound the change in potential during this serving phase in {\bf Case (a)} and {\bf Case (b)(ii)}. 
Note that if $e$ is the major edge, the server's final location is at a distance of at most $4\ell(e)$ 
from the server's initial solution 
(the server first reaches the major edge, then crosses the major edge, and eventually ends on the other side of some key edge). Therefore, the total change in potential is at most $4\kappa \ell(e)$, which is also bounded by $O(h^2)$ times the 
total length of the key edges.
\end{proof}

\section{Online Service with Delay: $k$ Servers}
\label{sec:kosd}

We now generalize the \osd problem to $k$ servers, for any $k \geq 1$. 
We call this the $k$-\osd problem. Again, we only give an algorithm for 
the $k$-\osd problem on HSTs, which yields an algorithm 
for the $k$-\osd problem on general metric spaces by a low-distortion 
tree embedding~\cite{FakcharoenpholRT04}. But, before stating this result, 
let us first note that the $k$-\osd problem generalizes the well-studied 
$k$-server problem~\cite{ManasseMS90,chrobak1991optimal,chrobak1991new,
KoutsoupiasP95,Koutsoupias99,BartalK04,BansalBMN15}. This latter problem is exactly identical 
to the $k$-\osd problem, except that every request must be served 
immediately on arrival
(consequently, there is no delay and therefore, no delay penalty).
But, this can be modeled in the \osd problem by using an infinite delay 
penalty for every request. 
Furthermore, this also establishes a connection between $k$-\osd 
and paging problems~\cite{SleatorT85,FiatKLMSY91,McGeochS91,Young94,AchlioptasCN00,Irani02,BansalBN12}, as will be discussed in Section~\ref{sec:uniform} and \ref{sec:star}.

We now re-state our results for $k$-\osd (Theorem~\ref{thm:kosd}
in the introduction).

\begin{theorem}
	There is a deterministic algorithm with a competitive ratio of $O(k h^4)$
    for the $k$-\osd problem on an HST of depth $h$.
    As immediate corollaries, this yields the following:
    \begin{itemize}
		\item A randomized algorithm with a competitive ratio of $O(k\log^5 n)$
	    for the $k$-\osd problem on general metric spaces.
        \item A deterministic algorithm with a competitive ratio of 
        $O(k)$ for the unweighted and weighted paging problems with delay 
        penalties, which are respectively special cases of $k$-\osd on 
        the uniform metric and star metrics.
   \end{itemize}
\end{theorem}	
	
We will discuss the paging problems in Section~\ref{sec:uniform} and
Section~\ref{sec:star}. In this section, we will focus on proving Theorem~\ref{thm:kosd}
on HSTs.

To generalize our algorithm for \osd from a single server to $k$ servers, 
we use ideas from the so-called {\em active cover} algorithm for $k$-server 
on trees by Chrobak and Larmore~\cite{chrobak1991optimal}. This algorithm 
is a generalization of the well-known {double cover} 
algorithm~\cite{chrobak1991new} for $k$-server on the line metric.


In the active server algorithm, we define {\bf active} servers as the ones that are directly connected to the request location on the tree, i.e., there is no other server on the path connecting the request to an active server. To serve a request, the algorithm moves {\em all} active servers simultaneously at the same speed to the request, until one of these servers reaches the request and serves it. 

\subsection{Algorithm}

We use the intuition of moving all active servers toward a request in our $k$-\osd algorithm as well. To describe the algorithm, we specify how we modify the waiting phase and serving phase of the algorithm.

\smallskip\noindent
{\bf Waiting Phase.} During the waiting phase, for each server, the algorithm has a separate counter on each edge and increments this counter exactly as in the \osd algorithm assuming this is the only server. The waiting phase ends when for any of these servers, a major edge is saturated. (Note that set of major edges may be different for different servers, since it depends on the location of the server.)

\smallskip
\noindent
{\em Remark:} In our $k$-\osd algorithm, we allow a server to reside in the middle of an edge during a waiting phase. If this is the case, and if for any request, the segment of the edge on which the server is residing is longer than all the other edges on the path between the server and the request, then we call this segment the major edge. Also, note that we do not need to actually place a server at the middle of an edge in an implementation; instead, we can simply delay moving the server until it reaches the other end of the edge. The abstraction of placing a server on an edge, however, will make it easier to visualize our algorithm, and to do the analysis.

\smallskip\noindent
{\bf Serving Phase.} 
We apply the same {\bf Time-Forwarding} subroutine from the \ps algorithm for the \osd problem to determine the key edges, as well as the edges that will be traversed in the relevant subtree. Now, the algorithm needs to decide which server to use in this service. This is decided in two phases. First, we order the key edges in a DFS order, with the property that siblings are ordered in non-decreasing length. We then assume that every key edge has a request at its end, and use the active cover algorithm to serve these (imaginary) requests. Initially, the active cover algorithm moves all active servers toward the first key edge, and the unique server that reaches it first actually crosses this key edge. We now need to serve an actual set of requests in the subtree below this key edge. To serve these requests, the server that traversed the key edge performs a DFS tour and returns to the base of the key edge. Note that we are not using the active cover algorithm during this service, i.e., all other servers do not move. Once this server returns to the base of the key edge, the active cover algorithm again resumes toward the next key edge in our pre-determined order. 
%
%
Once all requests that were selected to be served have indeed been served, the server that was serving requests below the last key edge returns to the base of this key edge.

\subsection{Analysis}

To analyze the new algorithm, we need to use the same potential function 
as the active cover algorithm~\cite{chrobak1991optimal}. Define potential function $\Phi$ to be 
$$\Phi = \kappa (k ||M_{\min}|| + \sum_{i<j}||s_is_j||).$$  
Here, $M_{\min}$ is the minimum-cost matching (\mcm) between the algorithm's servers and \opt's servers
for a fixed optimal solution \opt. $||M_{\min}||$ denotes the cost of the matching, i.e., the sum of lengths of edges in 
$M_{\min}$. In the potential function, $s_1, \ldots, s_k$ denote the locations of the algorithm's servers. 
We choose then parameter $\kappa$ to be $Ch^2$ for a large enough constant $C$ (note that
this is similar to what we did for the \ps algorithm with $k=1$). 

We first summarize the properties of this potential function.
(The subtree below a node $v$, including $v$ itself, is denoted by $T_v$.)

\begin{lemma}\label{lem:potential}
The first property concerns \opt's movement:
\begin{itemize}
	\item When \opt moves a server by distance $d$, the 
    potential function can increase by at most $\kappa k d$.
\end{itemize}

Now, suppose the algorithm moves all active servers at equal speed toward a node $v$ in the tree. Further, suppose the algorithm does not have any server in the subtree $T_v$. 
Then, we have:
\begin{itemize}\itemsep=0pt
\item If \opt has a server in $T_v$, then the potential function decreases by $\kappa$ times the total distance traveled by the algorithm's servers.
\item If \opt does not have a server in $T_v$, then the potential function can increase by at most $2k\kappa$ times the movement of any one of algorithm's servers. 
\end{itemize}
\end{lemma}
\begin{proof} 
We prove each property separately:
	\begin{itemize}
    \item By the triangle inequality, the cost of the previous \mcm 
    increases by at most $d$, and the cost of this matching is an 
    upper bound on the cost of the new \mcm.
    \item Suppose each of algorithm's servers travels distance $d$.
    Furthermore, let the number of active servers be $q$, where the 
    $i$th active server resides on the path connecting $r_i$ servers
    to $v$. Then, the distance between the $i$th active server and 
    $r_i$ servers increases by $d$, its distance to the 
    remaining $(k-r_i-q)$ inactive servers decreases by $d$,
    and its distance to the remaining $q-1$ active servers 
    decreases by $2d$. Overall, the change in the second term
    in $\Phi$ is given by:
    $$d \cdot \kappa \cdot \sum_{i=1}^q (2r_i-k+1) 
    = d \cdot \kappa \cdot \left[2 (k-q) - kq + q \right]$$
    $$= d \cdot \kappa \cdot \left[2 k-q - kq \right]  $$
    The change in the first term is only due to the $q$ active servers,
    of which one can wlog be matched with \opt's server in $T_v$.
    Therefore, the change in the first term is at most:
    $$d \cdot \kappa \cdot \left[k q-2 k\right].$$
    Adding these two expressions shows that the decrease of potential is 
    at least $\kappa \cdot (qd)$. 
    \item We again get the same expression as above for the change in the
    second term of $\Phi$. For the first term, the change is at most
    $$d \cdot \kappa \cdot k q.$$
    Adding these two expressions, the potential increases by at most
    $$\kappa \cdot (2kd - qd) \leq \kappa \cdot (2kd).\qedhere$$
	\end{itemize}
\end{proof}


Using these properties, we are now going to establish that
our $k$-\osd algorithm is $O(k h^3)$-competitive 
(Theorem~\ref{thm:kosd}).
%
%
We follow a similar strategy as our \osd algorithm --- we will again show the moving cost and waiting cost can all be charged to the cost of the key edges (we will lose an extra factor of $k$ here). There are several new issues that now arise that are specific to $k$ being more than $1$. But first, we show that service inside the subtrees $X_e$ for key edges $e$ is analogous to the single server scenario. (Recall the definition of $X_e$ from Section~\ref{sec:single-server-algo}.)

\begin{lemma} \label{noservers}
When the serving phase begins, none of the algorithm's servers is in the relevant subtree.
\end{lemma}

\begin{proof}
Suppose not: there is a server $s$ in the relevant subtree. Let $s'$ be the server whose major edge was saturated first. We will call edges saturated/unsaturated according to the state of their counters for $s'$. Consider the path between $s$ and the end of the major edge that is not in the relevant subtree $R_e$ (let us call this vertex $v$). (If the major edge is only part of an edge, then this path also ends at the same location in the middle of the edge. Indeed, throughout this lemma, we will treat the major edge as a full edge, although it may really only be a segment of an edge.) Traversing this path from $v$, let $e$ be the first edge that is unsaturated, and $e'$ be the saturated edge immediately before $e$ on this path. Note that $e'$ must exist because the major edge is saturated.

The length of edge $e$ is at most the length of $e'$ because either $e'$ is the parent of $e$ in the HST or $e'$ is the major edge. Let $u$ be the vertex shared by $e$ and $e'$. If there is a saturated edge $e''$ in the subtree $T_u$ that has length at least the length of $e$, then $e''$ is a saturated major edge for $s$. Otherwise, since edge $e'$ is saturated for $s'$, and all the requests in $X_{e'}$ will contribute to edge $e$ for server $s$, edge $e$ would be saturated before $e'$. Therefore, there cannot be a server in the relevant subtree.
%
\end{proof}

Note that Lemma~\ref{waitingcosts} still holds for $k$-\osd. Therefore, we can
charge all delay penalties to service costs of the algorithm up to a constant factor.

\begin{lemma}
\label{lma:delay-k}
	The total delay penalties incurred by the $k$-\osd algorithm is at most 
    the total service cost of the algorithm.
\end{lemma}

The service costs come in two parts. The first part is the cost of traversing edges
inside the subtrees $X_e$ for key edges $e$. Lemma~\ref{keyedge} still holds as well;
therefore, the service cost inside subtrees $X_e$ for key edges $e$ is at most 
$O(h^2)$ times the sum of lengths of key edges. 

\begin{lemma}
\label{lma:service-k}
	The total service cost inside the subtree $X_e$  for a key edge $e$ is at most
    $O(h^2)$ times $\ell(e)$, the length of $e$.
\end{lemma}

Now, we need to
analyze the service cost of traversing the subtree connecting the major edge and the 
key edges, including both the major edge and the key edges. These are the parts 
where the active cover algorithm is being used.

First, we state a general property analogous to Lemma~\ref{clm:potentialdecrease},
that bounds the change in potential when the servers move.

\begin{lemma}
\label{lma:potential-k}
At any time in the active cover algorithm, if all of the algorithm's servers are moving 
toward some server in \opt, then the potential decreases by $\kappa$ times the 
distance moved by the servers. 
\end{lemma}
\begin{proof}
This follows directly from Lemma~\ref{lem:potential}. 
\end{proof}

Furthermore, let us consider a subtree $X_e$ for a key edge $e$, where $X_e$ does not 
contain any of \opt's servers. In this case, we use the same charging tree argument as 
in the \ps algorithm. Note that since Lemma~\ref{lma:history} still holds. So,
we can apply Lemmas~\ref{lem:chargingtree} and \ref{lma:opt-chargingtree} to bound 
the lengths of the key edges against \opt's costs up to a factor of $O(h)$.

\begin{lemma}
\label{lma:chargingtree-k}
	The sum of lengths of all key edges $e$ such that the subtrees $X_e$ does not contain 
    any of \opt's servers is at most $O(h)$ times the cost of \opt.
\end{lemma}
\begin{proof}
	This follows directly from Lemmas~\ref{lem:chargingtree} and \ref{lma:opt-chargingtree}.
\end{proof}

First, note that if the major edge is the solitary key edge, then
Lemmas~\ref{lma:service-k}, \ref{lma:potential-k}, and \ref{lma:chargingtree-k} 
are sufficient to bound the total service cost by $O(kh^3)$ times the cost
of \opt. Therefore, in the rest of this section, we will assume that the key 
edges are distinct from the major edge.

\smallskip\noindent
{\em
For a saturated edge $e$ that is between the major edge and key edges 
(including the major edge and key edges themselves), we call the subtree 
$X_e$ a {\bf charging branch} if:
\begin{itemize}\itemsep0pt
\item 
\opt does not have a server in $X_e$, and 
\item 
$e$ is the major edge, or \opt has a server in the subtree $X_{p(e)}$ 
for the parent $p(e)$ of $e$ in the critical tree $C_e$.
\end{itemize}
In this case, we call $e$ the {\em root} edge of the charging branch.
}

\smallskip\noindent
Note that every key edge $e$ such that \opt does not have a server in $X_e$ 
belongs to one of the charging branches. Conversely, any charging branch 
contains one or more such key edges. The total service cost of the 
algorithm in a charging branch is defined as the algorithm's cost between 
serving the last key edge outside charging branch, and serving the last key 
edge inside the charging branch.

\begin{lemma}\label{lem:keysubtree}
The sum of algorithm's service cost in serving requests in a charging branch
and the change in potential is at most $O(kh^3)$ times the sum of lengths of 
key edges in the charging branch.
\end{lemma}

\begin{proof}
Let $e$ be the root edge of the charging branch. The service cost of the algorithm in the charging branch 
can be divided into three parts:
\begin{enumerate}\itemsep0pt
	\item the service cost incurred between serving the last request outside the charging branch 
    till a server reaches edge $e$,
    \item the service cost of using the active cover algorithm between the key edges in $S$, and
    \item the cost of preemptive service below the key edges.
\end{enumerate}

Let $\ell$ be the sum of lengths of key edges in the charging branch.

Lemma~\ref{lma:service-k} implies that the service cost of (3) is at most $O(h^2 \ell)$. 
Next, observe that by the definition of key edges, the service cost of (2) for any single 
server is at most $O(h \ell)$, and therefore, at most $O(kh\ell)$ counting all the servers.
Furthermore, by Lemma~\ref{lem:potential}, the increase in potential due to (2) (note that
(3) does not change potential) is at most 
$2k\kappa$ times the service cost of any single server, the latter being at most $O(h \ell)$. 
Therefore, the total service cost and increase in potential due to (2) and (3)
is at most $O(kh^3 \ell)$.

Now we deal with cost of (1). If $e$ is the major edge, 
then the distance traversed by any single server to the major edge 
is $O(1)$ times the length of the major edge, which is $O(\ell)$
since the sum of lengths of key edges is at least the length of the 
major edge. The corresponding change in potential is $O(kh^2\ell)$.

Now, suppose $e$ is not the major edge. In this case, note that 
$p(e)$, the parent 
edge of $e$, must be in the critical tree $C_e$ and that OPT is in $X_{p(e)}$. Then, there are two cases:
\begin{itemize}
	\item 
    First, consider the situation that the algorithm has not served any request
    in the current serving phase before reaching edge $e$. In this case, 
    algorithm's servers are moving toward one of \opt's servers until one of them 
    traverses $e$. By Lemma~\ref{lma:potential-k},
    the decrease in potential is sufficient to account for the service cost 
    of the algorithm in (1). 
    \item
    Next, consider the situation that the algorithm did serve requests
    in the current serving phase before reaching edge $e$. Let $e'$ be the 
    last key edge for which the algorithm served requests in $X_{e'}$.
    There are two subcases:
    \begin{itemize}
    	\item 
        Suppose $p(e)$, the parent edge of $e$ in the critical subtree $C_e$, is 
        not an ancestor of $e'$. Then, we know that the algorithm does not 
        have a server in $X_{p(e)}$ by Lemma \ref{noservers} and the 
        fact that no request in $X_{p(e)}$ has been served yet. In this case,
        we again have the property that 
        algorithm's servers are moving toward one of \opt's servers until one of them 
	    traverses $e$. By Lemma \ref{lma:potential-k},
    	the decrease in potential is sufficient to account for the service cost 
	    of the algorithm in (1). 
        \item
        Now, suppose $p(e)$ is an ancestor of $e'$ as well. Recall that in the DFS order
        used by the algorithm between key edges, siblings are ordered in non-decreasing
        length. This implies that the distance from $e'$ to $e$ is at most $2$ times the
        length of $e$. As a consequence, the total distance traversed by a single server
      	in (1) is at most $2$ times the length of $e$. Therefore, we can charge the 
        service cost in (1) to $O(\ell)$ and the potential increase in (1) to
        $O(kh^2 \ell)$.\qedhere
    \end{itemize}
\end{itemize}

%
%
\end{proof}

We are now ready to bound the total cost of the algorithm against the cost of \opt.

\begin{lemma}
\label{lma:kosd-main}
	The total service and delay cost in the $k$-\osd algorithm 
    is $O(kh^4)$ times the cost of \opt.
\end{lemma}
\begin{proof}
First, we bound the total service cost.
We partition the total movement of any single server into phases: we say that that 
the algorithm is in phase $e$ for a key edge $e$ is the active cover algorithm is
moving the active servers toward the imaginary request at the end of edge $e$, 
or a server is serving requests in $X_e$. 

Phases are of two kinds:
\begin{itemize}
	\item 
    If \opt has a server in $X_e$, then the sum of potential change and service
    costs in phase $e$ is non-positive by Lemma~\ref{lma:potential-k}.
    \item
    We are left with the phases where \opt does not have a server in $X_e$ for 
    phase $e$. These phases are partitioned into super-phases defined by the 
    charging branch that $e$ belongs to - that is, for each charging phase we define a super-phase, and then include phase $e$ in the super-phase of the charging branch $e$ belongs to. For each such super-phase, by 
    Lemma~\ref{lem:keysubtree}, the total service cost and potential change
    is at most $O(kh^3)$ times the length of the key edges in the 
    charging branch, $\ell$. Furthermore, by Lemma~\ref{lma:chargingtree-k}, 
    $\ell$ can be bounded by $O(h)$ times the cost of $\opt$.
\end{itemize}    
Since the potential function is non-negative, it follows that the total 
service cost of the $k$-\osd algorithm is at most $O(kh^4)$ times the cost of \opt,

Finally, by Lemma~\ref{lma:delay-k}, we bound the total delay penalties of the 
algorithm to be at most the total service cost. This completes the proof.
\end{proof}

This shows that the competitive ratio of the $k$-\osd algorithm is $O(kh^4)$.
\section{Special Metrics}
\label{sec:special}
In this section, we present algorithms for the $k$-\osd
problem on special metric spaces: uniform and star metrics.
These respectively correspond to the paging and weighted
paging problems with delay.

\subsection{Uniform Metric/Paging}
\label{sec:uniform}

We give a reduction from the $k$-\osd problem on a uniform metric
to the classical online paging problem. The reduction is online, 
non-clairvoyant, and preserves the competitive ratio up to a constant.
To describe this reduction, it would be convenient to view the 
$k$-\osd problem on a uniform metric as the following equivalent
problem that we call {\em paging with delay}:

\smallskip\noindent
{\em There are $n$ pages that issue requests over time. Each request
is served when the corresponding page is brought into a cache 
that can only hold $k < n$ pages at a time. The algorithm can 
bring a page into the cache and evict an existing page, this 
process being called a {\em page swap} and having a cost of one.
For every request, the algorithm also incurs a delay penalty,
which is a monotone function of the delay in serving the 
request. The algorithm can only access the current value of the delay penalty.}

\smallskip\noindent
Classical paging is the special case of this problem
when every request has an infinite delay penalty, i.e., each
request must be served immediately, and the total cost is 
the number of page swaps. 

\begin{figure}
	\centering
	\begin{subfigure}[b]{0.4\textwidth} 
    	\includegraphics[width=\columnwidth]{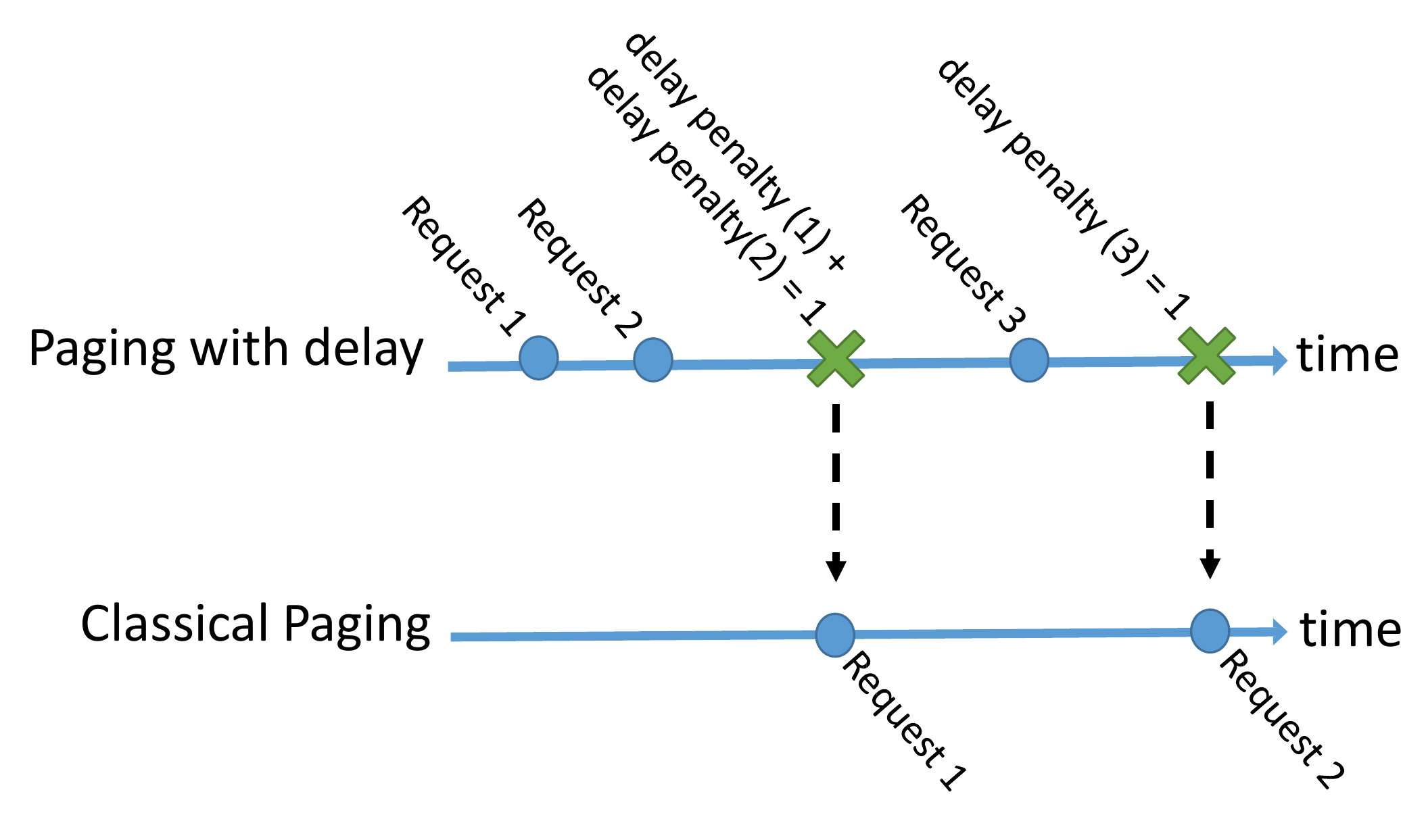}
		\caption{The reduction between instances for unweighted paging. 
        	This figure shows the requests for a single page.}
		\label{fig:reduction-upper}
	\end{subfigure}
	\begin{subfigure}[b]{0.4\textwidth}
        	\includegraphics[width=\columnwidth]{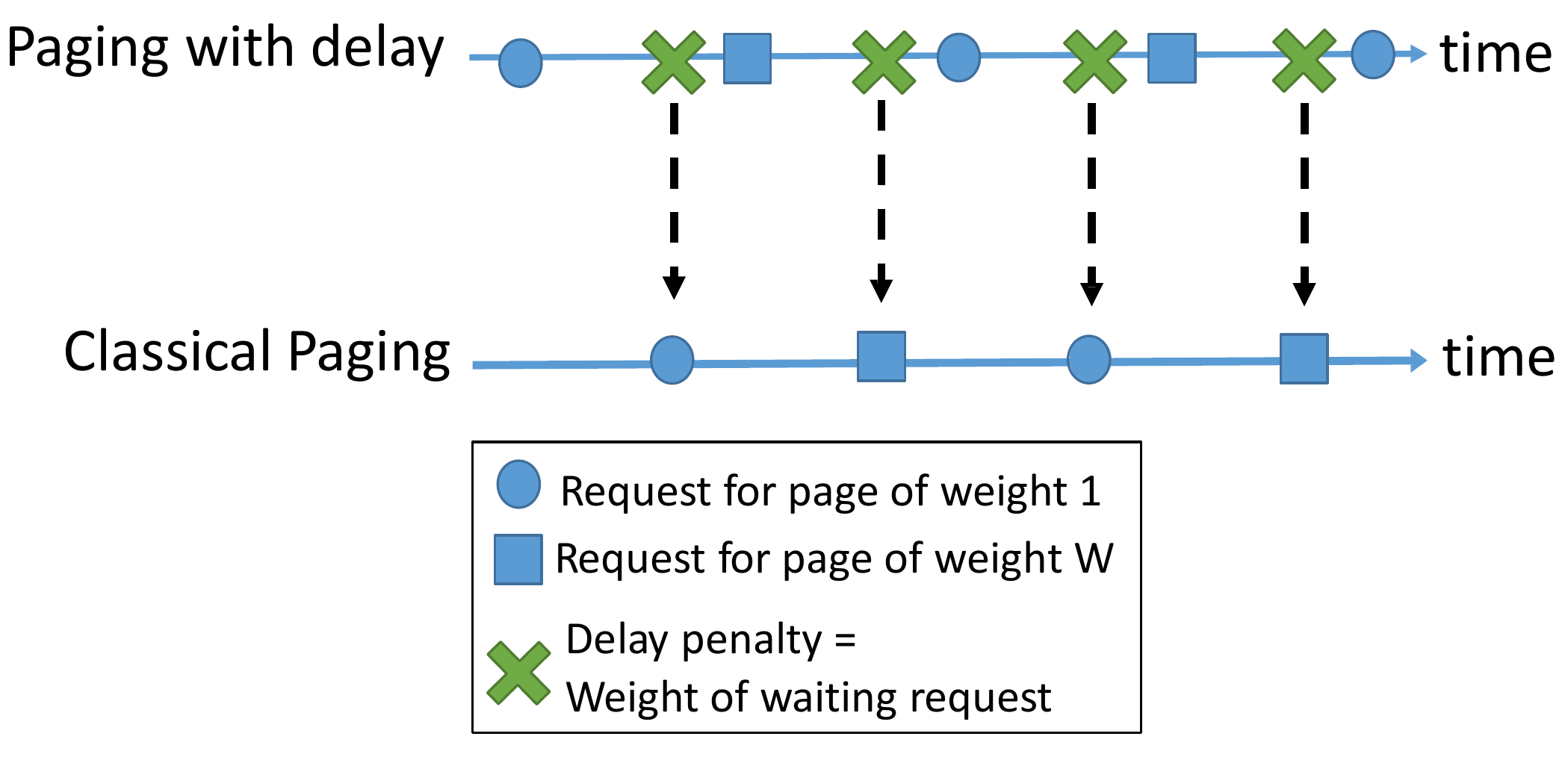}
		\caption{An example where the reduction has a large gap for weighted paging.}
		\label{fig:reduction-lower}
	\end{subfigure}
	\caption{\small The reduction between instances of online paging with delay and classical 
	online paging: it works for unweighted paging, but fails for weighted paging.}
\end{figure}

Suppose we are given an instance $\mathcal{I}$ of the paging problem 
with delay. We will reduce it to an instance $\mathcal{I'}$ of 
classical online paging. The requests for a page $p$ in $\mathcal{I}$ 
are partitioned into time intervals such that the total delay penalty 
incurred by the requests in an interval is exactly one at the end 
of the interval. In instance $\mathcal{I'}$, this interval is now replaced 
by a single request for $p$ at the end of the interval. 
(See Fig.~\ref{fig:reduction-upper}.) 
Note that the reduction is online and non-clairvoyant. 
We now show that by this reduction, {\em any} algorithm for 
$\mathcal{I'}$ can be used for $\mathcal{I}$, losing only a constant in 
the competitive ratio.  

\begin{lemma}
\label{lma:reduction}
	The optimal cost in $\mathcal I'$ is at most 3 times the optimal cost in $\mathcal I$. Conversely,
	the cost of an algorithm in $\mathcal I$ is at most 2 times that in $\mathcal I'$.
\end{lemma}
\begin{proof}
	Let us denote the optimal costs in $\mathcal I$ and $\mathcal I'$ by $\opt_{\mathcal I}$ and $\opt_{\mathcal I'}$
	respectively. Similarly, let $\alg_{\mathcal I}$ and $\alg_{\mathcal I'}$ respectively denote the 
	cost of an algorithm in $\mathcal I$ and $\mathcal I'$.

	Construct a solution for $\mathcal I'$ that maintains exactly the same cache contents as
	the optimal solution for $\mathcal I$ at all times. Whenever there is a request for a page 
	that is not in the cache, it serves the page and then immediately restores the cache
	contents of $\opt_{\mathcal I}$ by performing the reverse swap and restoring the previous 
	page in the cache. Note that there are two types of page swaps being performed:
	the first category are those that are also performed by $\opt_{\mathcal I}$, and the second
	category are the pairs of swaps to serve requests for pages not in the cache. The total 
	number of page swaps in the first category is identical the number of page swaps in 
	$\opt_{\mathcal I}$. For the second category, if a page $p$ is not in the cache at the end 
	of an interval in $\mathcal I$, there are two possibilities. If no request for $p$ in the
	preceding interval was served in $\opt_{\mathcal I}$ during the interval, then the 
	total delay penalty for these requests at the end of the interval is 1. On the other hand, 
	if some request for $p$ in the preceding interval was served in $\opt_{\mathcal I}$ during the
	interval, but the page is no longer in the cache at the end of the interval, then there 
	was at least 1 page swap that removed $p$ from the cache during the interval. 
	In either case, we can charge the two swaps performed in $\mathcal I'$ 
	 to the cost of 1 that $\opt_{\mathcal I}$ incurs in this interval for page $p$.
	
	In the converse direction, note that by definition, the swap cost of $\alg_{\mathcal I}$
	is identical to that of $\alg_{\mathcal I'}$. The only difference is that in $\mathcal I$,
	the algorithm also suffers a delay penalty. The delay penalty for a page in an interval
	is 0 if the algorithm maintains the page in the cache throughout the interval,
	and at most 1 otherwise. In the latter case, the algorithm must have swapped the page 
	out during the interval since the page was in the cache at the end of the previous 
	interval. In this case, we can charge the delay penalty in $\alg_{\mathcal I}$ to this 
	page swap during the interval that $\alg_{\mathcal I'}$ incurred.
\end{proof}

Using the above lemma and the well-known and asymptotically tight 
$O(k)$-competitive deterministic and 
$O(\log k)$-competitive randomized algorithms for online paging 
(see, e.g., \cite{BorodinE98}), we obtain the following theorem.

\begin{theorem}
\label{thm:uniform}
	There are $O(k)$-competitive deterministic and $O(\log k)$-competitive 
    randomized algorithms for online paging with delay (equivalently, 
    $k$-\osd on the uniform metric). This implies, in particular, that 
    there is an $O(1)$-competitive deterministic algorithm for 
    \osd on uniform metrics. These results also apply to the non-clairvoyant
    version of the problems, where the delay penalties in the future are 
    unknown to the algorithm. These competitive ratios are asymptotically 
    tight.
\end{theorem}

\subsection{Star Metric/Weighted Paging}
\label{sec:star}

As in the case of the uniform metric, it will be convenient to view 
the $k$-\osd problem on a star metric as a paging problem with delay,
the only difference being that every page $p$ now has a non-negative 
weight $w_p$ which is the cost of evicting the page from the cache. Note that in this case the aspect ratio $\Delta$ of the star metric is (up to a constant factor) 
the ratio of the maximum and minimum page weights.

For weighted paging, suppose we try to use the reduction strategy 
for (unweighted) paging from the previous section. 
We can now partition requests for
a page $p$ into intervals, where the total delay penalty at the end 
of the interval is the weight of the page $w_p$. This reduction, 
however, fails even for simple instances. Consider two pages of weights $W$ and $1$, where 
$W \gg 1$. Suppose their requests have penalty equal to their respective weights for unit delay.
If the requests for the pages alternate, and the cache can only hold only a single page, the 
algorithm repeatedly swaps the pages, whereas an optimal solution keeps the heavy page in the
cache and serves the light page only once every $W$ time units. The gap induced by the 
reduction in this instance is $W$. (See Fig.~\ref{fig:reduction-lower}.)

Indeed, we show that this difficulty is fundamental in that there is no
good non-clairvoyant algorithm for the weighted paging problem with delay. 

\begin{theorem}
\label{thm:non-clairvoyant}
	There is a lower bound of $\Omega(W)$ on the competitive ratio of weighted paging 
    with delay, even with a single-page cache (equivalently, \osd on a star metric) 
    in the non-clairvoyant setting, where $W$ is the ratio of the maximum to minimum 
    page weight (equivalently, aspect ratio $\Delta$ of the star metric).
\end{theorem}
\begin{proof}
Suppose there is a heavy page $p_0$ with cost $W$ (without loss of generality assume $W$ is an integer), and all other pages $p_1, p_2, ..., p_n$ have cost $1$. Set $n = W^2$. We will construct an adversarial instance based on how the algorithm decides to serve the pages. The instance is constructed in phases: at the beginning of each phase, there is a request for all of the small pages $p_1,...,p_n$. However, the waiting cost for all of these requests start at $0$, and at some time $t$ (not necessarily in the same phase) will become infinity; we say that this request becomes critical at time $t$. The adversary will control the time that requests become critical. For each phase, we maintain the following properties:
\begin{enumerate}
\item At the beginning of each phase, all the small pages have a request.
\item The algorithm either serves all the small pages at least once, or serves page $p_0$ at least $W$ times, in each phase.
\item The optimal solution does not incur any waiting cost. In each phase, the optimal solution serves $p_0$ once and at most $W$ of the small pages.
\end{enumerate}
Given these properties, it is easy to check that the competitive ratio of the algorithm is at least $W/2$, because in each phase algorithm spends at least $W^2$, while the optimal spends at most $2W$. 

Now we describe how the adversary works. At time $\epsilon$ (for any fixed $\epsilon$ strictly between $0$ and $1$), there is a request for $p_0$ that requires immediate service. At time $t = 1,2,...,W$, if the algorithm has not served all the small pages at least once in this phase, let $r_t\in \{1,2,...,n\}$ denote the index of an unserved page. The adversary makes $p_{r_t}$ critical at time $t$, and adds a request for $p_0$ at time $t+\epsilon$. Note that page $p_0$ must be in the algorithm's cache at time 
$t - 1 + \epsilon$ and at time $t+\epsilon$, but cannot be in the algorithm's cache at time $t$. Thus, the algorithm must evict $p_0$ at least once in this interval. Therefore, at the end of time $W+1$, either the algorithm has served all the small pages at least once in this phase, or it has served $p_0$ at least $W$ times.

The optimal solution serves $p_{r_1}$, $p_{r_2}$, $\dots$, $p_{r_t}$ at time $0$, and $p_0$ at time $\epsilon$. Since no other page became critical in this phase, the optimal solution will never need to evict $p_0$ in this phase. 

After time $W+1$, the adversary generates a request for each of the small pages, and starts a new phase. If this procedure repeats for $k$ phases, then the algorithm's cost is at least $kW^2$, while the optimal solution's cost is at most $2kW + n$ (the optimal solution serves all non-critical requests at the very end). Therefore, as $k$ goes to infinity, the competitive ratio of the algorithm cannot be better than $W/2$.
\end{proof}

The above theorem implies that we must restrict ourselves to the clairvoyant
setting for the weighted paging with delay problem. For this case, 
our results for \osd/$k$-\osd on a star metric follow 
as corollaries of our theorems for general 
metrics (Theorems~\ref{thm:osd-hst}, \ref{thm:kosd}).

\begin{theorem}
\label{thm:star}
	There is an $O(k)$-competitive deterministic algorithm for online 
    weighted paging with delay (equivalently, $k$-\osd on a star metric). 
    This implies, in particular, an $O(1)$-competitive deterministic algorithm 
    for the \osd problem on star metrics. These competitive ratios are 
    asymptotically tight for deterministic algorithms.
\end{theorem}

\section*{Acknowledgement}
Y. Azar was supported in part by the Israel Science Foundation (grant No. 1506/16), 
by the I-CORE program (Center No.4/11) and by the Blavatnik Fund. A. Ganesh 
and D. Panigrahi were supported in part by NSF Awards CCF 1535972 and CCF 1527084. 
This work was done in part while Y. Azar and D. Panigrahi were visiting the
Simons Insitute for Theoretical Computer Science, as part of the 
{\em Algorithms and Uncertainty} program.


\bibliographystyle{plain}
\bibliography{ref,server}

\end{document}